\documentclass[12pt,a4paper]{article}
\usepackage[utf8]{inputenc}

\usepackage[utf8]{inputenc}
\usepackage{graphicx} 
\usepackage{caption,subcaption}
\usepackage{amsmath}
\graphicspath{{images/}} 

\begin{document}

\begin{center}

{\Large {\bf 
A Model of Mass Extinction Accounting for Species's Differential Evolutionary Response to a Catastrophic Climate Change
 }

\vspace*{11mm}

Amer Alsulami$^{a,b}$ and Sergei Petrovskii$^a$\footnote{Corresponding author. Email: sp237@le.ac.uk}
 }

\vspace*{9mm}

$^a~$ School of Computing and Mathematical Sciences, University of Leicester, Leicester, LE1 7RH, UK\\

\vspace*{2mm}

$^b$ Taraba University College, Taif University, Saudi Arabia

\end{center}

\vspace{7mm}

\begin{center}
{\bf Abstract}
\end{center}

Mass extinction is a phenomenon in the history of life on Earth when a considerable number of species go extinct over a relatively short period of time. The magnitude of extinction varies between the events, the most well known are the ``Big Five'' when more than one half of all species got extinct. There were many extinctions with a smaller magnitude too. It is widely believed that the common trigger leading to a mass extinction is a climate change such a global warming or global cooling. There are, however, many open questions with regard to the effect and potential importance of specific factors and processes. In this paper, we develop a novel mathematical model that takes into account two factors largely overlooked in the mass extinctions literature, namely, (i) the active feedback of phytoplankton to the climate through changing the albedo of the ocean surface and (ii) the species's adaptive evolutionary response to a climate change. We show that whether species goes extinct or not depends on a subtle interplay between the scale of the climate change and the rate of the evolutionary response. We also show that species's response to a fast climate change can exhibit long transient dynamics (false extinction) when the species population density remain at a low value for a long time before recovering to its safe steady state value. Finally, we show that the distribution of extinction frequencies predicted by our model is generally consistent with the fossil record. 

\vspace{10mm}


\vspace{9mm}


\newpage

\section{Introduction}

Mass extinction is a phenomenon that several times punctuated the functioning of global biota during 541 Myrs (the Phanerozoic Eon) of the recorded history of the life on Earth \cite{alroy2008,bambach2006,sepkoski1986}. 
Species get extinct all the time at some background rate (eventually being replaced by new, emerging species); this is regarded as a normal course of macroevolution \cite{jablonski1986,stanley1979,wang2003}. However, sometimes the species extinction rate exceeded the average rate by more than an order of magnitude resulting in disappearing, over a relatively short period of time, of 50-75\% of all the existing species \cite{bambach2006,bond2017}: a mass extinction occurred. 

Mass extinctions have been attracting a considerable and ever-growing scientific attention over the last several decades \cite{alvarez1980,hallam1997,harnik2012,newman2003,raup1982,sole1997}. It is now widely believed that the main trigger leading to a mass extinction is a sufficiently large (overcritical) perturbation of the CO$_2$ cycle~\cite{rothman2017,rothman2019a,rothman2019b}. This can occur for different reasons, e.g.~a massive volcano eruption called a LIP (Large Igneous Province) \cite{bond2017} or a collision of Earth with a large bolide \cite{alvarez1980}. 
An overcritical perturbation of the CO$_2$ cycle then leads to species extinction through a variety of mechanisms or pathways, e.g.~global warming or cooling, ocean acidification and ocean anoxia \cite{harnik2012}.  

While a considerable progress in understanding mass extinctions has been made over the last two decades (e.g.~see the references above), yet many questions and issues remain \cite{sudakow2022}. One potentially important issue is the effect of the active feedback of a vegetation - in particular, phytoplankton - to a climate change. Studies on mass extinctions tend to regard species as passive entities that merely accept an environmental change (and get extinct if the climate change makes the environment too harsh). However, this is not always true as, in fact, some taxa can fight back to transform the environment according to their needs. On small temporal and spatial scales, such species are called ``ecosystem engineers'' \cite{hastings2007,jones1994}, on a global scale this effect is summarized by the Gaia concept \cite{lenton2018,lovelock2016,lovelock1974}. 
For instance, phytoplankton, when present in high densities, is known to be have a feedback on climate \cite{charlson87}, however it remains unclear whether such effect can be extended to a global scale, e.g.~to accelerate or slow down a mass extinction. 

Another open question is the role of species phenotypic plasticity: species can adapt to an environmental change, at least to a certain extent \cite{pigliucci2001,price2003}. It seems obvious that plasticity can slow down the extinction rates through a directed change of species traits, hence affecting the extinction magnitude; however, this issue has been largely unaddressed in the literature. 

The inherent deficiency of the fossil record, see \cite{sudakow2022} for details, severely impedes further progress in understanding mass extinctions, in particular in distinguishing between the effects of different factors and mechanisms. This is a situation where, arguably, mathematical modelling can be used to partially compensate for the shortcomings of the data \cite{sudakow2022}. Surprisingly, the challenges of understanding mass extinctions have been largely overlooked by applied mathematicians. 
The literature concerned with mass extinctions modelling mostly focuses on a statistical analysis of the data, e.g.~revealing the power laws in the frequency distribution \cite{newman1997,newman2003,sole1997} or attempting to reveal a pattern in the extinction timing \cite{melott2014,raup1984}. 
There are very few papers attempting at a dynamical modelling of mass extinctions \cite{chiba1998,feulner2011,roopnarine2006,sznajd2001}. 
We mention here that there is a considerable literature dealing with modelling of extinctions of individual species (e.g.~\cite{keith2008,lande1987,valenti2004} but there is a huge epistemological gap between extinction of a particular species at a particular location and a mass extinction that can wipe out more than 50\% of species globally. Upscaling the local/specific processes to a global scale is a highly nontrivial problem. 

In this paper, we develop a novel model that attempts to describe mass extinctions by considering an interplay between an active feedback of vegetation (particularly, phytoplankton) to a climate change and a differential species evolutionary response to the corresponding environmental change.  
The paper is organised as follows. In Section \ref{nu_1}, we develop the baseline single-species model that includes the phytoplankton feedback on the global energy balance and the dependence of the plankton growth rate on the temperature (but not an evolutionary response). In Section \ref{sec:bifurc}, we investigate the properties of the baseline model to reveal possible species extinctions resulting from bifurcations of the steady states. In Section \ref{sec:evolut1sp}, the baseline model is extended to include species's adaptive evolutionary response to a climate change. We show that extinction or persistence of the species depends on the interplay between the magnitude of the climate change and the evolutionary rate. We also show that the model exhibits long transient behaviour (false extinctions), so that species recovery from small densities can take a very long time. In Section \ref{nu_2}, we turn the single species model to a multi-species one. We investigate how the extinction frequencies depend on the values of the key parameters or their distributions and in Section \ref{sec:compar} 
we endeavour to tentatively compare the simulation results with the fossil record on mass extinctions. Finally, in Section \ref{sec:discuss} we discuss our findings and outline possible directions of future work.

\section{Global conceptual climate-vegetation model}\label{nu_1}

Following a widely used conceptual approach~\cite{fan2021,sellers,lohmann2020,soldatenko2021}, in this study we assume that the state of the Earth climate can be described by a single variable, i.e.~the average Earth temperature. Correspondingly, for the dynamics of the global climate we use the zero-dimensional Budyko-Sellers model described by the following equation \cite{sellers}:
\begin{equation}\label{sellers}
\lambda\frac{dT}{dt} = -e \sigma T^4 + \frac{\mu_0 I_0}{4}(1- S),
\end{equation}
where $T$ is the average surface temperature, $t$ is time and $\lambda$ is a coefficient known as thermal inertia. In the right-hand side of Eq.~(\ref{sellers}), the first and second terms describe, respectively, the outgoing light emission (described by the Stefan-Boltzmann law with the effective emissivity $e$) and the fraction of the incoming solar radiation that falls on the Earth surface where $S$ is the surface albedo, $I_0$ quantifies the total amount of solar energy and $\mu_0$ is a coefficient. 

\begin{figure}[!t]
\centering
\includegraphics[scale=0.9]{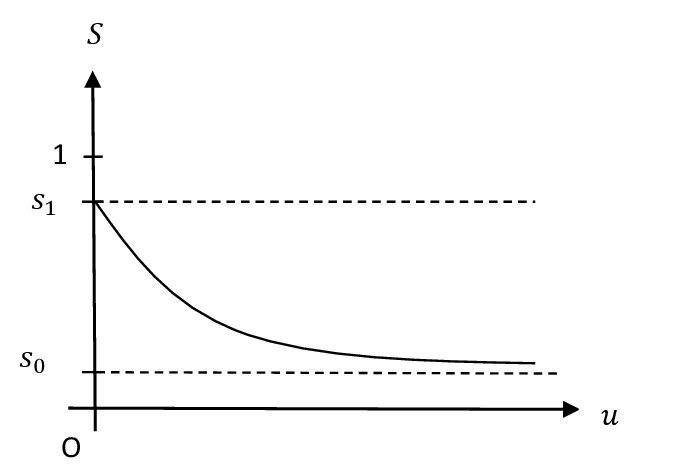}\hspace{0mm}\includegraphics[scale=0.85]{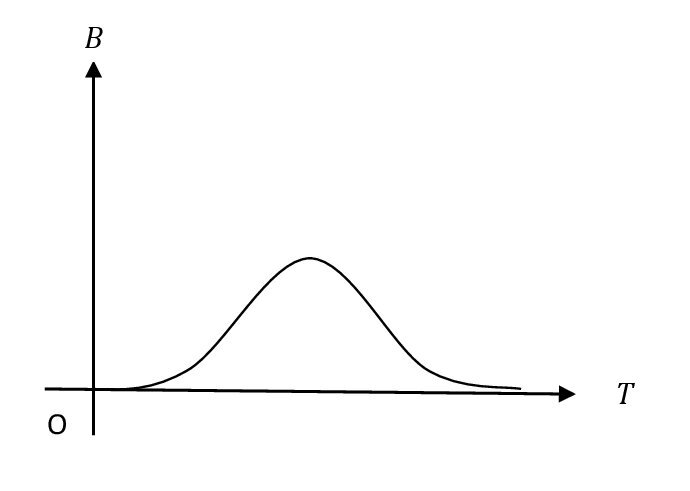}\\
\hspace{0mm} (a) \hspace{65mm} (b)
\caption{{\small (a) Generic dependence of the ocean surface albedo $S$ on phytoplankton density $u$. (b) Generic dependence of the phytoplankton per capita growth rate $B$ on the water temperature $T$.}}
\label{s0s1} 
\end{figure}

Obviously, any change in the Earth surface albedo will change the balance between the incoming and outgoing energy fluxes, thus resulting in an increase or decrease of the average temperature and hence potentially leading to a global climate change. One factor that may change the albedo is known to be vegetation cover. While the effect of the terrestrial vegetation is still under-investigated, the effect of the aquatic vegetation - i.e.~phytoplankton - is well established~\cite{charlson87}: the presence of phytoplankton tends to decrease the ocean surface albedo; see Fig.~\ref{s0s1}a. Since about 70\% of Earth is covered by oceans, any global tendency in phytoplankton density to increase or decrease is likely to have a significant effect on the Earth temperature. 

For our baseline model, we assume that, as well as the temperature, the ocean phytoplankton can also be described by a single variable, i.e.~the average phytoplankton density in the upper ocean layer - say, $u(t)$ - hence considering phytoplankton as a single `meta-species'. Apparently, this disregards not only the geographical variations in the plankton density but also the presence of different plankton species; the effect of the latter will be discussed below. Arguably, in such schematic, conceptual model the dynamics of phytoplankton as a whole can be modelled similarly to the dynamics of individual populations, e.g.~by considering the logistic growth~\cite{haney1996,heyman1983,steel1992}:
\begin{equation}\label{phytmeta}
\frac{du}{dt} = \left[B(T)-\frac{u}{K}\right]u-\sigma u,
\end{equation}
where $\sigma$ is the mortality rate and $K$ is the carrying capacity, i.e.~the maximum amount of phytoplankton that can be sustainably contained by a unit volume of the sea water. Note that the phytoplankton reproduction rate $B(T)$ depends on the water temperature, for which there is considerable empirical evidence. It is well known that the growth rate increases for low and intermediate temperatures,in particular in the range $0<T<30^{\circ}$~\cite{kremer2017}. The temperature dependence for higher water temperatures (e.g.~for $T>40^{\circ}$) is less known but it seems intuitively obvious that it should be decreasing (as, ultimately, too high temperatures will kill phytoplankton altogether). Thus, if considered over a sufficiently broad temperature range, function $B(T)$ should be bell-shaped (see Fig.~\ref{s0s1}b). 

In order to complete the model, we now need a specific parametrization of functions $S(u)$ and $B(T)$. While it is clear that the dependence of the surface albedo on the phytoplankton density is a monotonously decreasing function (cf.~Fig.~\ref{s0s1}a), its precise shape is unknown. Since our goal here is to build a model which is qualitative rather than quantitative (i.e.~which predicts the tendencies but not necessarily specific numbers), instead of looking for a precise shape, we use the following hypothetical function:
\begin{equation}\label{albedo}
S(u) = (s_1-s_0)e^{-\alpha_1 u}+s_0, 
\end{equation}
where $s_0$, $s_1$ and $\alpha_1$ are coefficients. Note that, while the exponential decay in (\ref{albedo}) is hypothetical, the marginal values $S(0)$ and $S(\infty)$ (and hence coefficients of $s_0$ and $s_1$) can be estimated based on available data for the value of the ocean surface albedo for a clean water and for the turbid water with high plankton density, 

For the dependence of the phytoplankton growth rate on the water temperature, we use the following parametrization: 
\begin{equation}\label{plankgrowtemp}
B(T)=b_1e^{-\frac{T_0}{T}}e^{-\alpha_2 T}, 
\end{equation}
where $b_1$, $T_0$ and $\alpha_2$ are coefficients. It is readily seen that $B(T)$ defined by function (\ref{plankgrowtemp}) has a unique maximum at $T_m=(T_0/\alpha_2)^{1/2}$ and decreases to zero for $|T-T_m|\gg 1$. 
Note that the first exponent in (\ref{plankgrowtemp}) is essentially the Arrhenius law describing the generic dependence of chemical reactions (i.e., in our case, metabolic reactions) on the temperature, cf.~\cite{kremer2017}. 

Thus, we arrive at the following temperature-phytoplankton system:
\begin{align}
     \frac{dT}{dt}&=\frac{1}{\lambda}\left(-aT^4+b[1-(s_1-s_0)e^{-\alpha_1 u}-s_0]\right),\label{sys2T}\\
     \frac{du}{dt}&=[b_1e^{-\frac{T_0}{T}}e^{-\alpha_2 T}-u]u-\sigma u,\label{sys2u}
\end{align}
where we have introduced coefficients $a=e\sigma$ and $b=\frac{\mu_0 I_0}{4}$ for the notations simplicity.

\section{Bifurcations and extinctions in the conceptual model}\label{sec:bifurc}

	\begin{figure}[!b] 
		\begin{subfigure}[b]{0.5\linewidth}
			\centering
			\includegraphics[width=1\linewidth]{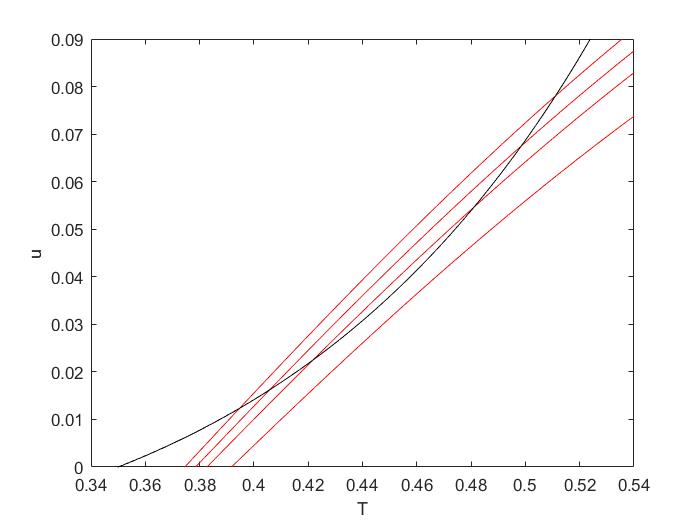} 
			\caption{} 
			\label{1b1} 
			\vspace{4ex}
		\end{subfigure}
		\begin{subfigure}[b]{0.5\linewidth}
			\centering
			\includegraphics[width=1\linewidth]{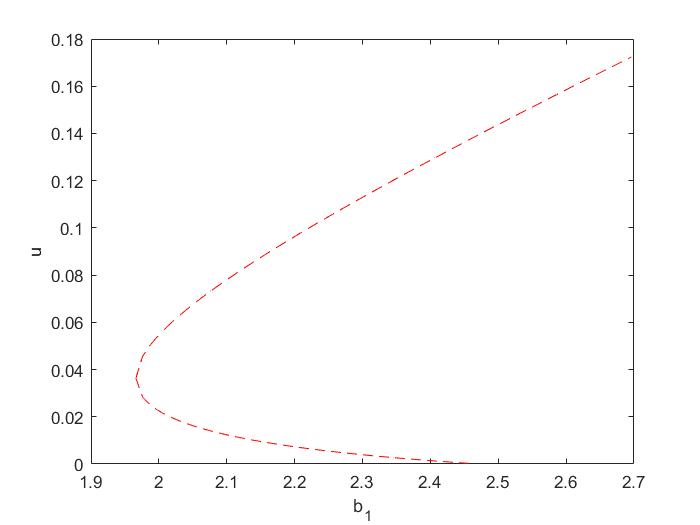} 
			\caption{} 
			\label{2b1} 
			\vspace{4ex}
		\end{subfigure} 
		\begin{subfigure}[b]{0.5\linewidth}
			\centering
			\includegraphics[width=1\linewidth]{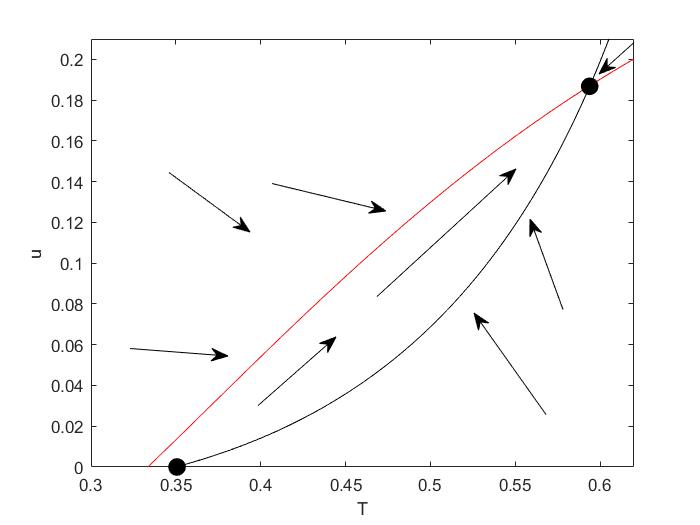} 
			\caption{} 
			\label{3b1} 
		\end{subfigure}
		\begin{subfigure}[b]{0.5\linewidth}
			\centering
			\includegraphics[width=1\linewidth]{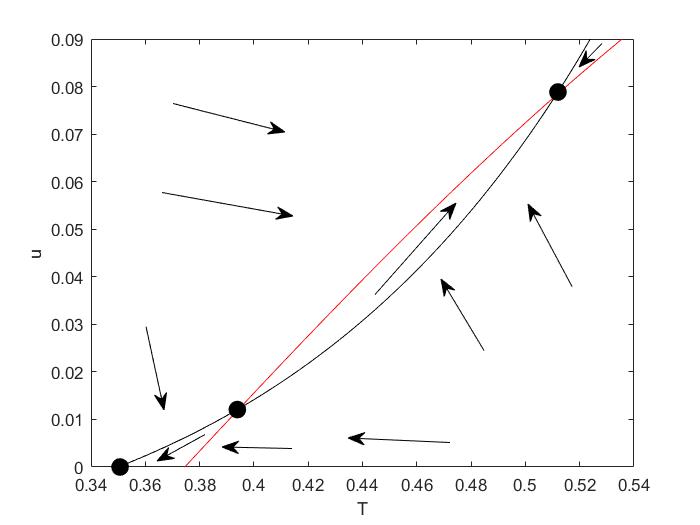} 
			\caption{}
			\label{4b1} 
		\end{subfigure} 
\caption{{\small (a) The (null) isoclines of the temperature–phytoplankton system (\ref{=T}-\ref{=u}). Black curve shows the ﬁrst (temperature) isocline (\ref{=T}) obtained for $a=1,\ b=0.3,\ s_0=0.1,\ s_1=0.95$, and $\alpha_1=3$; red curves show the second (phytoplankton) isocline (\ref{=u})  obtained for $b_1=2.1,\ 2.05,\ 2$, and $ 1.9$ (from left to right, respectively), $\alpha_2=1,\ T_0=1$, and $\sigma=0.1$. (b) The steady state values of $u$ (obtained as the intersection points of the two isoclines) as a function of the controlling parameter $b_1$. (c) and (d)  A sketch of the phase plane of the temperature–phytoplankton system obtained for parameters $b_1=2.8$, and $b_1=2.1$, respectively, other parameters as in (a). The black and red curves show the temperature isocline and phytoplankton isocline, respectively, large black dots show the steady states, black arrows show sample trajectories of the system and the direction of the phase ﬂow as given by vector $(\frac{dT}{dt},\ \frac{du}{dt})$.}}
		\label{fig1.3}
	\end{figure}

We now investigate the properties of the model (\ref{sys2T}-\ref{sys2u}), particularly the number and stability of its steady states and how that may depend on the model's parameters. 
The steady states of the dynamical system (\ref{sys2T}-\ref{sys2u}) are the solutions of the following algebraic system:
	\begin{align}
		\frac{1}{\lambda}[-aT^4+b(1-(s_1-s_0)e^{-\alpha_1 u}-s_0)]&=0,\label{=T}\\
		[b_1e^{-\frac{T_0}{T}}e^{-\alpha_2 T}-u]u-\sigma u&=0.\label{=u}
	\end{align}
Equations (\ref{=T}-\ref{=u}) deﬁne the two (null) isoclines of the system, which we will call the temperature isocline and the phytoplankton isocline, respectively. The temperature isocline is therefore given by
	\begin{equation}\label{isoc1}
		u = {} -\frac{1}{\alpha_1} \ln{\left(\frac{\frac{a T^4}{b}+s_0-1}{s_0-s_1}\right)}.
	\end{equation}
The phytoplankton isocline consists of two parts, i.e., of the following curve
	\begin{equation}\label{isoc2}
		u=b_1e^{-T \alpha_2}e^{-\frac{T_0}{T}}-\sigma,
	\end{equation}
and the straight line $u=0$.

It does not seem possible to solve the system (\ref{=T}-\ref{=u}) analytically. However, given the fact the system's equilibria are the intersection points of the two isoclines, it is possible to reveal the tendencies leading to the system's bifurcations (e.g.~the disappearance or emergence of the steady states) by revealing the change of the relative position of the isoclines (i.e.~the relative position of the corresponding curves) in response to a change in parameter values. For the reasons that will be explained in Section \ref{sec:evolut1sp}, we consider $b_1$ as the main controlling parameter but we also reveal the effect of some other parameters such as $s_0$, $s_1$ and $\sigma$. 

Figure \ref{1b1} shows a few cases of possible relative position of the isoclines (\ref{isoc1}) and (\ref{isoc2}) for different values of parameter $b_1$. It is readily seen that the positive equilibria are only feasible if $b_1$ is above a certain critical value. This is summarized in the bifurcation diagram shown in Fig.~\ref{2b1}. 
For sufficiently large values of $b_1$, the system (\ref{=T}-\ref{=u}) has an unstable semi-trivial equilibrium $(\bar{T_0},0)$ and a stable positive equilibrium $(\bar{T_1},\bar{u_1})$; see Fig.~\ref{3b1}. 
When $b_1$ falls below a certain critical value, a transcritical bifurcation happens so that  the semi-trivial equilibrium becomes stable and another positive equilibrium (a saddle) appears, say $(\bar{T_2},\bar{u_2})$; see Fig.~\ref{4b1}. Along with a further decrease in $b_1$, the two positive steady states move toward each other, so that when it falls below the second critical value, the positive steady states merge and disappear: a saddle-node bifurcation happens. Thus, for sufficiently small values of $b_1$, the system possesses only the semi-trivial steady state $(\bar{T_0},0)$.

	\begin{figure}[!t] 
		\begin{subfigure}[b]{0.5\linewidth}
			\centering
			\includegraphics[width=1\linewidth]{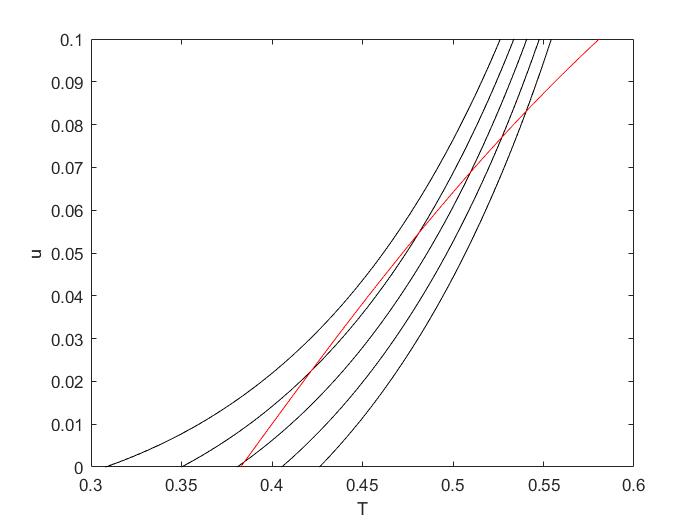} 
			\caption{} 
			\label{1s1} 
			\vspace{4ex}
		\end{subfigure}
		\begin{subfigure}[b]{0.5\linewidth}
			\centering
			\includegraphics[width=1\linewidth]{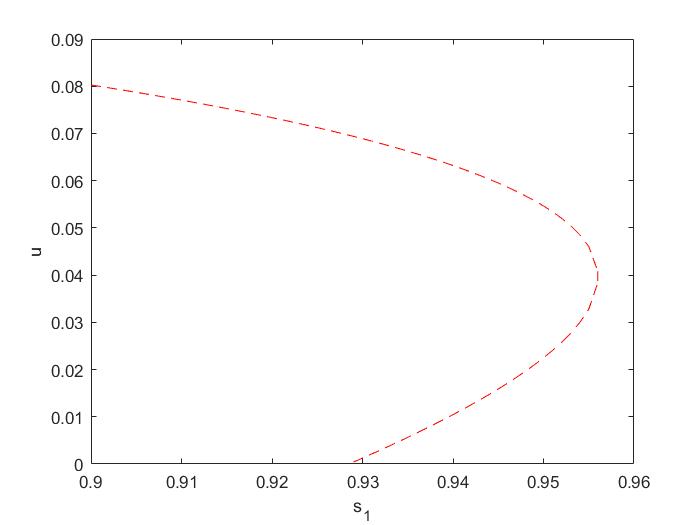} 
			\caption{} 
			\label{2s1} 
			\vspace{4ex}
		\end{subfigure} 
		\begin{subfigure}[b]{0.5\linewidth}
			\centering
			\includegraphics[width=1\linewidth]{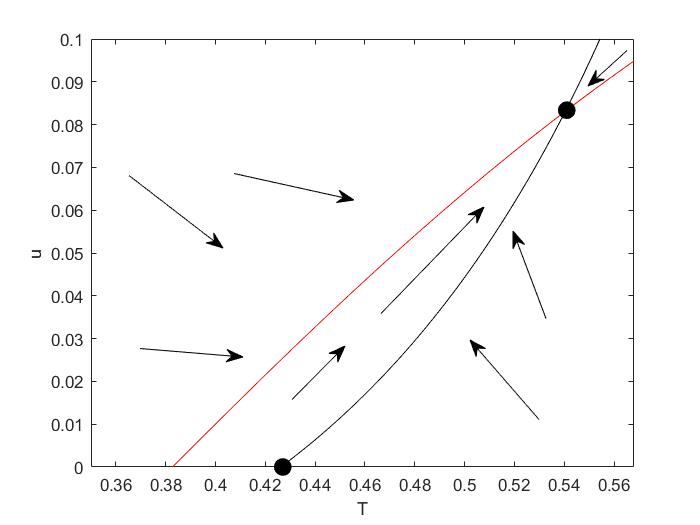} 
			\caption{} 
			\label{3s1} 
		\end{subfigure}
		\begin{subfigure}[b]{0.5\linewidth}
			\centering
			\includegraphics[width=1\linewidth]{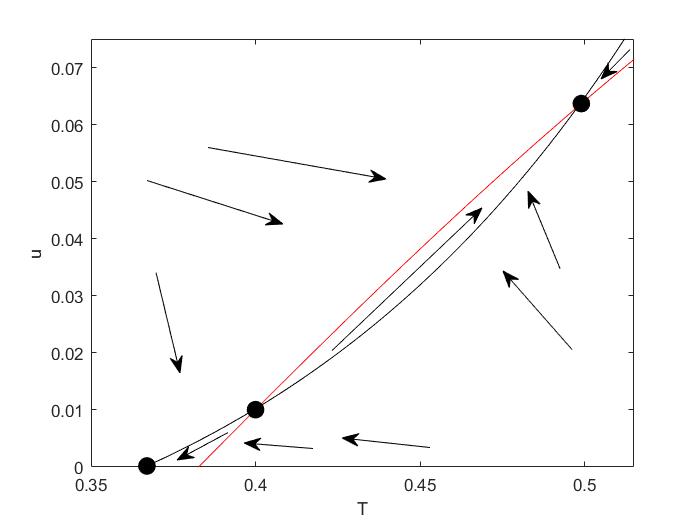} 
			\caption{}
			\label{4s1} 
		\end{subfigure} 
\caption{{\small (a) The (null) isoclines of the temperature–phytoplankton system (\ref{=T}-\ref{=u}). Black curves show the ﬁrst (temperature) isocline (\ref{=T}) obtained for $s_1= 0.97,\ 0.95,\ 0.93,\ 0.91$, and $ 0.89$ (from left to right, respectively), $a=1,\ b=0.3,\ s_0=0.1$, and $\alpha_1=3$; red curve shows the second (phytoplankton) isocline (\ref{=u}) obtained for $b_1=2,\  \alpha_2=1,\ T_0=1$, and $\sigma=0.1$. (b) The steady state values of $u$ (obtained as the intersection points of the two isoclines) as a function of the controlling parameter $s_1$. (c) and (d) A sketch of the phase plane of the temperature–phytoplankton system obtained for parameters $s_1=0.89$, and $s_1=0.94$, respectively, other parameters as in (a). The black and red curves show the temperature isocline and phytoplankton isocline, respectively, large black dots show the steady states, black arrows show sample trajectories of the system and the direction of the phase ﬂow as given by vector $(\frac{dT}{dt},\ \frac{du}{dt})$.}} 
		\label{fig7} 
	\end{figure}

For the stability of the steady states (which can be be established straightforwardly by considering the direction of the phase flow, see the arrows in Figs.~\ref{3b1} and \ref{4b1}), we readily see that in the  case where the system has a semi-trivial equilibrium and one positive equilibrium (cf.~Fig.~\ref{3b1}) the positive state is a stable node, while the semi-trivial state is a saddle (hence unstable). For the case, when we have one semi-trivial steady state and two positive steady states (cf.~Fig.~\ref{4b1}), it is clear to see that the semi-trivial state and the upper positive state are stable nodes, while the lower positive state is a saddle (hence unstable). 

We now consider $s_1$ as the controlling parameter. 
The change in the relative position of the isoclines and the corresponding bifurcation structure of the system are shown in Fig.~\ref{fig7}. It is readily seen that a similar succession of changes occurs in response to an increase in $s_1$, so that the positive equilibria are only feasible if $s_1$ is small, i.e.~below a certain critical value.
For sufficiently small values of $s_1$, the system (\ref{=T}-\ref{=u}) possesses an unstable semi-trivial equilibrium $(\bar{T_0},0)$ and a stable positive equilibrium $(\bar{T_1},\bar{u_1})$. When $s_1$ increases to a certain value, a transcritical bifurcation happens so that $(\bar{T_0},0)$ becomes stable a saddle-point $(\bar{T_2},\bar{u_2})$ emerges. When $s_1$ increases to the second critical value, the two positive steady states move toward each other, so that they eventually merge and disappear; see Fig.~\ref{2s1}.

In the case where either $s_0$ or $\sigma$ are considered as controlling parameters, the system shows a sequence of qualitative changes in its properties similar to that shown in Fig.~\ref{fig7}; we therefore do not show those results here for the sake of brevity. 

\begin{figure}[ht]
		\centering
		\begin{subfigure}{.4\linewidth}
			\includegraphics[width = \linewidth]{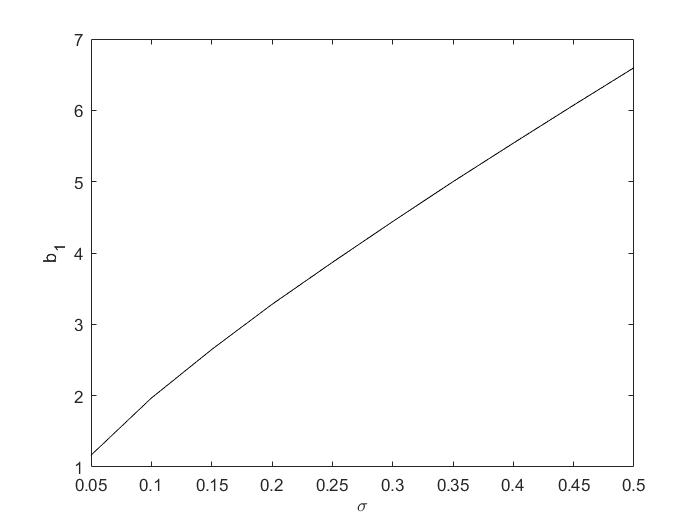}
			\caption{}
			\label{analysis_1}
		\end{subfigure}%
		\hspace{1em}
		\begin{subfigure}{.4\linewidth}
			\includegraphics[width = \linewidth]{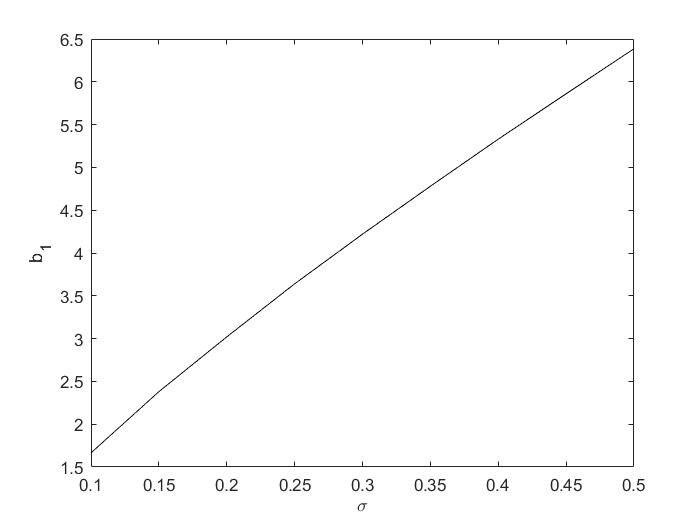}
			\caption{}
			\label{analysis_6}
		\end{subfigure}
\caption{{\small The saddle-node bifurcation value of $b_1$ for different values of $\sigma$ in the system (\ref{sys2T}-\ref{sys2u}) obtained for $s_1=0.95\ \text{and}\ 0.9$, in (a) and (b) respectively. Other parameters are given in the text. For the parameters from below the curve, the system possesses only the semi-trivial equilibrium $(\bar{T_0},0)$ (which is a stable node in this parameter range).}}
		\label{analysis}
	\end{figure}

Note that the bifurcation values of parameter $b_1$ are not universal but depend on the values of other parameters, in particular on $\sigma$. As an example, Figure (\ref{analysis}) shows the critical values of $b_1$ corresponding to the saddle-node bifurcation in the system (\ref{sys2T}-\ref{sys2u}) as a function of $\sigma$ obtained for two different values of $s_1$ and having other parameters fixed at the same hypothetical values as above ($\lambda=1,\ a=1,\ b=0.3,\ s_0=0.1,\ \alpha_1=3,\ \alpha_2=1,\ \text{and}\ T_0=1$). 

\begin{figure}[h] 
\centering
\includegraphics[width=0.32\linewidth]{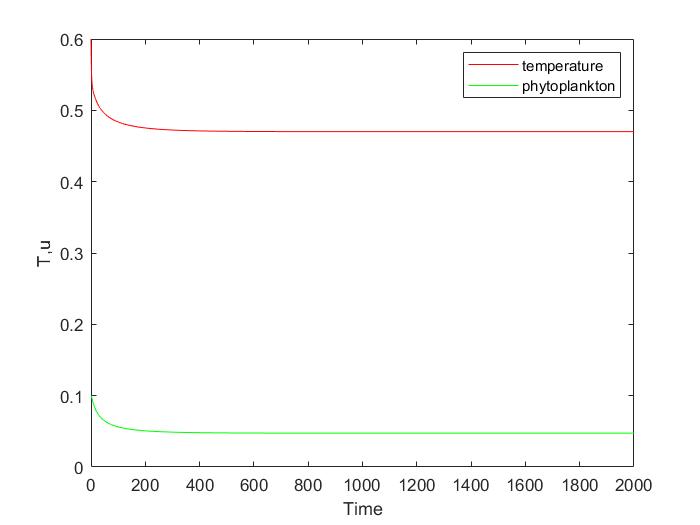} 
\includegraphics[width=0.32\linewidth]{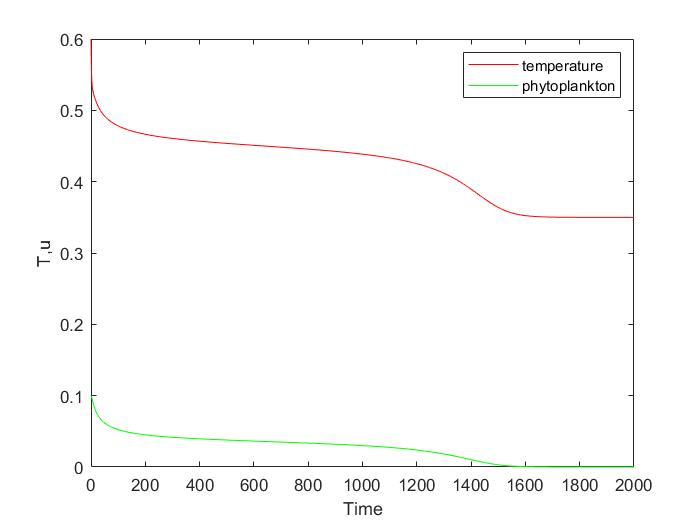} 
\includegraphics[width=0.32\linewidth]{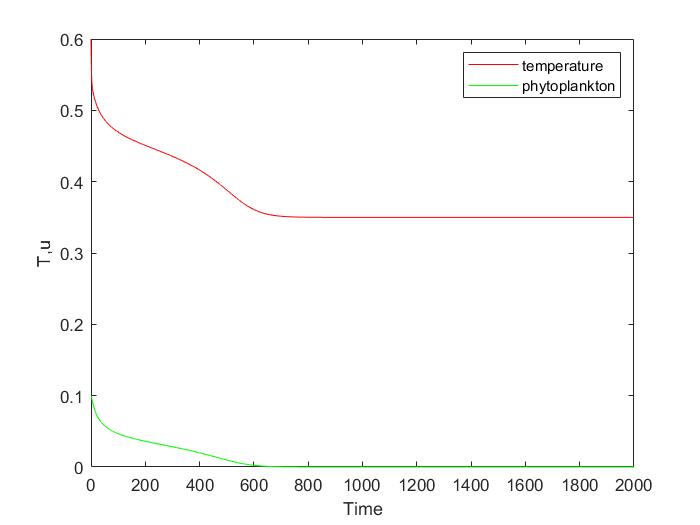} \\
\hspace{9mm} (a) \hspace{45mm} (b) \hspace{45mm} (c)
\caption{{\small Averaged temperature (red) and phytoplankton's density (green) versus time obtained for the initial conditions $T_0=0.6$, $u_0=0.1$, and parameter (a) $b_1=1.98$, (b) $b_1=1.96$ and (c) $b_1=1.93$. Other parameters are given in the text.}} 
  \label{fig1} 
\end{figure}

In order to obtain the temperature and the the phytoplankton density as functions of time, system (\ref{sys2T}--\ref{sys2u}) is solved numerically. Figure \ref{fig1} shows the results obtained for parameter values  $b=0.3,\ a=1, \ T_0=1, \ \alpha_2=1,\ \sigma=0.1, \ \alpha_1=3, \ s_0=0.1, \ s_1=0.95, \ \lambda=1$ and a few different values of $b_1$. In Fig.~\ref{fig1}a, we choose $b_1=1.98$ which is a subcritical value, i.e.~when the system possesses a stable positive equilibrium (see Fig.~\ref{2b1}). In this case, the initial conditions fast approach their asymptotical positive steady state values. Figures \ref{fig1}b and \ref{fig1}c are obtained for supercritical values $b_1=1.96$ and $b_1=1.93$, respectively, i.e.~after the saddle-node bifurcation occurs, so that there are no positive equilibria. Correspondingly, in the course of time the system eventually approaches the semi-trivial steady state $(\bar{T_0},0)$, hence resulting in the phytoplankton extinction. 

Interestingly, for the values of $b_1$ below the bifurcation value but close to it, the extinction does not actually happen until after a rather long period of time when the state variables change very slowly, cf.~Fig.~\ref{fig1}b. This type of dynamics is known as a `long transient' \cite{hastings2018,morozov2020} and, in our case, is a result of the ghost attractor \cite{izhikevich2007,strogatz1994,tyukin2011}. 
The stable equilibrium has disappeared in the saddle-node bifurcation but, in the part of the phase pane where it was before the bifurcation, the phase flow remain slow. Thus, the system spends a long time there, which, ultimately (for a parameter value sufficiently close to its bifurcation value), may give an impression of a steady state - a ghost attractor. For a smaller value of $b_1$, i.e.~further away from its bifurcation value, the long transient becomes shorter (cf.~Fig.~\ref{fig1}c) and eventually disappears.

\section{Single-species model with adaptive evolution}\label{sec:evolut1sp}

One important phenomenon that is entirely missed by the above analysis is possible species adaptation to the environmental changes. 
Indeed, in the model (\ref{sys2T}-\ref{sys2u}), all parameters are fixed and that means that the species traits remain unchanged in the course of the dynamics. 
Meanwhile, there is considerable evidence showing that, in many cases, populations can evolve sufficiently to survive in new or altered environments, e.g.~resulting from the climate change or human-related activities such as biological invasions, pollution, etc.~\cite{bradshaw1981,hoffmann1991,smith1989,peters1992}). 

Models of adaptive evolution are numerohoffmann1991,us and any overview of them is beyond the scope of this study. Here we use the approach based on the conceptual model by Gomulkiewicz \& Holt \cite{gomulkiewicz1995} who showed that a modification of a given species's trait in response to a sudden environmental change can be described by an exponential transition of the corresponding parameter from its `old' (before change) value to a new one. In our model, a change in the temperature affects the phytoplankton growth rate. We therefore assume that the phytoplanckton evolves with time in order to reach an equilibrium with the changed environment. Specifically, to account for this adaptive evolution we assume that parameter $b_1$ is the following function of time:
\begin{equation}
b_1(t) = b_{\text{new}}-(b_{\text{new}}-b_{\text{old}})e^{-\gamma t}, \label{b_1(t)}
\end{equation}
where $b_{\text{old}}=b_1(0)$, $b_{\text{new}}=b_1(t\rightarrow\infty)$ and $\gamma$ is the parameter quantifying the rate of adaptive evolution. 

\begin{figure}[!b]
  \centering
  \begin{subfigure}{.4\linewidth}
    \includegraphics[width = \linewidth]{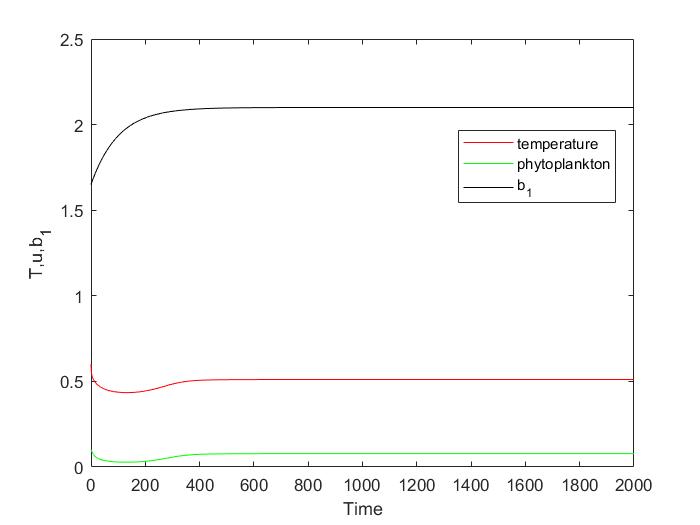}
    \caption{}
    \label{fig2_a}
  \end{subfigure}%
  \hspace{0em}
  \begin{subfigure}{.4\linewidth}
    \includegraphics[width = \linewidth]{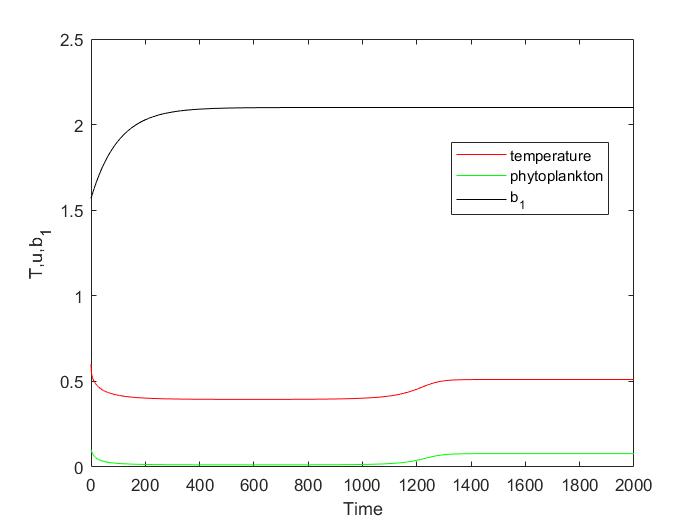}
    \caption{}
    \label{fig2_c}
  \end{subfigure}
    \hspace{0em}
  \begin{subfigure}{.4\linewidth}
    \includegraphics[width = \linewidth]{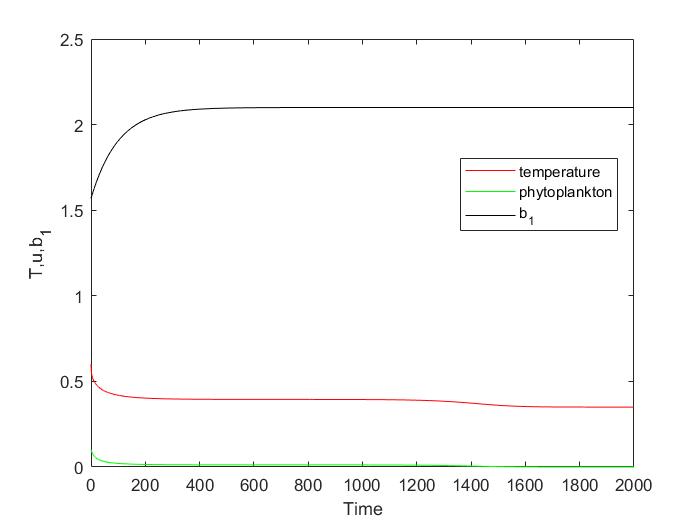}
    \caption{}
    \label{fig2_d}
  \end{subfigure}%
  \begin{subfigure}{.4\linewidth}
    \includegraphics[width = \linewidth]{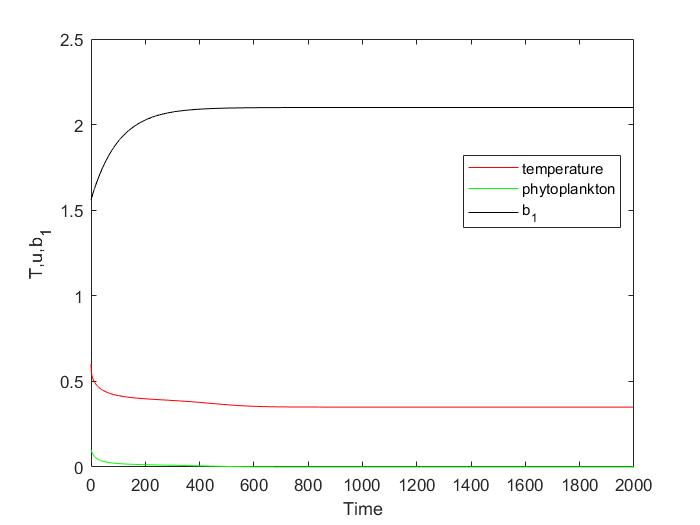}
    \caption{}
    \label{fig2_e}
  \end{subfigure}%
\caption{{\small Averaged temperature (red) and phytoplankton density (green) vs time as given by system (\ref{sys2T}-\ref{sys2u}) with $b_1(t)$ (black) defined by Eq.~(\ref{b_1(t)}) for different values of parameter $b_{\text{old}}$: (a) $b_{\text{old}}=1.65$, (b) $b_{\text{old}}=1.56874$, (c) $b_{\text{old}}=1.56873$ and (d) $b_{\text{old}}=1.56$. Other parameters are given in the text. The initial conditions are $T_0=0.6$, $u_0=0.1$.}}
  \label{fig2}
\end{figure}

Thus, $b_{\text{old}}$ is the parameter value that ensures that, before the climate change, the phytoplankton is at the positive steady state. The effect of a change in $b_1$ can be understood in terms of the bifurcation diagram shown in Fig.~\ref{2b1}. The equilibrium phytoplankton density is the point at the upper branch of the bifurcation curve corresponding to $b_1=b_{\text{old}}$. A climate change that we consider to be sufficiently fast suddenly pushes the bifurcation curve away. As a result, the plankton density is not at at equilibrium any more and hence undergoes the dynamics as prescribed by the Eqs.~(\ref{sys2T}-\ref{sys2u}). A big question is then whether this dynamics brings the phytoplankton to a new positive steady state (i.e.~another point at the upper steady branch of the bifurcation curve) or to extinction. 

Apparently, in terms of our approach the answer to the above question depends on two factors. The first factor is how far the sudden environmental change pushed the bifurcation curve away, i.e.~how far the before-change value of the phytoplankton density is away from the new position of the bifurcation curve. The second factor is the evolutionary rate: intuitively, a high enough evolutionary rate should help the species to avoid extinction. 

\begin{figure}[!b]
  \centering
    \includegraphics[width = 0.6\linewidth]{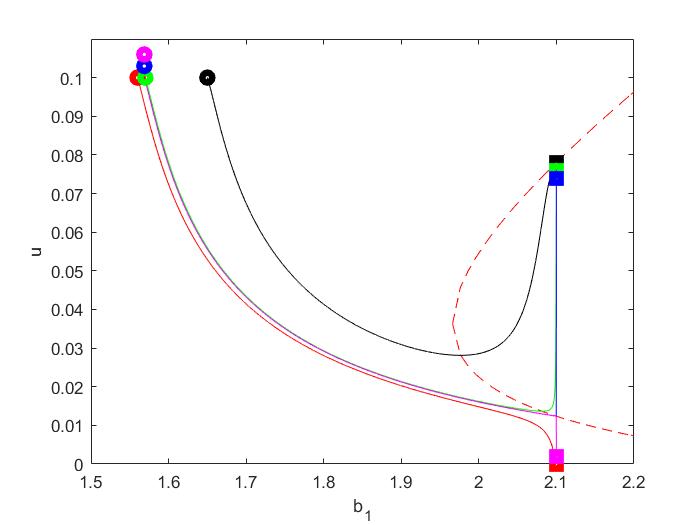}
\caption{{\small Dynamics of the system (\ref{sys2T}-\ref{sys2u},\ \ref{b_1(t)}) shown in the $(b_1,u)$ plane for $b_{\text{old}}=1.65$, 1.57, 1.56874, 1.56873 and 1.56 (black, green, blue, magenta and red colours, respectively); see details in the text. The circles are the start points and the squares are the end points. The dashed red curve shows the steady state values of $u$ (the bifurcation curve) in the baseline temperature-phytoplankton system (\ref{sys2T}-\ref{sys2u}) without evolution, cf.~Fig.~\ref{fig1.3}.}}
\label{traj} 
\end{figure}
\begin{figure}[!b]
\centering
\includegraphics[width=0.6\linewidth]{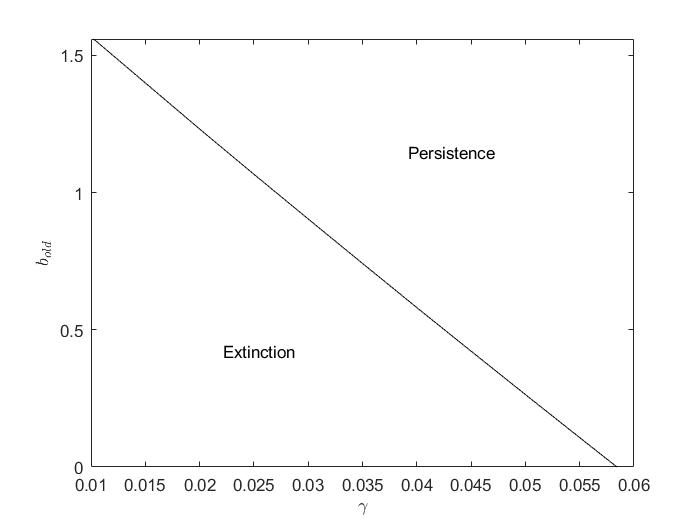}
 \caption{{\small The critical values of $b_{\text{old}}$ against $\gamma$ for system (\ref{sys2T}-\ref{sys2u}) with $b_1$ as a function of time. Parameter values are given in the text.}}
\label{fig3} 
\end{figure}

To illustrate the effect of the first factor, Fig.~\ref{fig2} shows the solution of the system (\ref{sys2T}-\ref{sys2u},\ \ref{b_1(t)}) obtained numerically for parameter values $b_{\text{new}}=2.1, \ \gamma=0.01$ and different values of $b_{\text{old}}$, other parameters are the same as above. (Note that, for the convenience of simulations, we quantify the difference between the pre-change and the post-change equilibrium values of parameter $b_1$ by varying $b_{\text{old}}$ rather than $b_{\text{new}}$.) 
We readily observe that there is a critical value of the difference $\delta=b_{\text{new}}-b_{\text{old}}$. For $\delta$ less than the critical value, the system overcomes the effect of the climate change so that, after an initial decay in the values of $T$ and $u$, they approach their new positive steady state values; see Figs.~\ref{fig2_a} and \ref{fig2_c}. However, for $\delta$ larger than the critical value, after a certain period of time when $T$ and $u$ show almost no changes, they eventually approach the semi-trivial steady state where phytoplankton goes to extinction and the temperature stabilizes at a lower value; see Figs.~\ref{fig2_d} and \ref{fig2_e}. 

Interestingly, as well as in our baseline model without species adaptation, for the values of $\delta$ close to its critical value, the final transition of the system to its asymptotic state, whether it is to the positive state or to the semi-trivial (extinction) state, does not happen until after a rather long period of a quasi-steady state dynamics when $T$ and $u$ do not show any visible changes (cf.~Figs.~\ref{fig2_c} and \ref{fig2_d}). This is another example of the long transient dynamics~\cite{hastings2018,morozov2020} shown by our global climate-vegetation model, cf.~Fig.~\ref{fig1}. 

Figure \ref{traj} visualizes the system's dynamics in a different way by showing the results of Fig.~\ref{fig2} as trajectories in $(b_1,u)$ plane (where black, green, blue, magenta and red curves correspond to $b_{\text{old}}=1.65, 1.57, 1.56874, 1.56873$ and $1.56$, respectively). 
Note that, in all cases, the value of $b_{\text{old}}$ is smaller than the critical value of $b_1$ (which corresponds to the left-most point of the bifurcation curve, cf.~Fig.~\ref{2b1}), so that without adaptation (i.e.~for $\gamma=0$, hence constant $b_1$) the phytoplankton extinction would happen inevitably in all cases. 
Under the effect of adaptive evolution ($\gamma=0.01>0$), we observe that some of the species survives (as shown by black, green and blue curves) in case the difference $\delta$ between $b_{\text{new}}$ and $b_{\text{old}}$ is sufficiently small (smaller than a certain critical value) to eventually settle down on a new equilibrium value of the phytoplankton density. In case $\delta$ is larger than the critical value, the species goes extinct, cf.~the magenta and red curves.

Note that, in the context of species response to a sudden climate change, parameter $b_{\text{old}}$ has a double interpretation. Firstly, different values of $b_{\text{old}}$ may be thought of as different species. Secondly, since the difference $\delta=b_{\text{new}}-b_{\text{old}}$ is a measure of the magnitude of the climate change, for a fixed $b_{\text{new}}$ parameter $b_{\text{old}}$ can also be regarded as a measure of the magnitude of the climate change. Namely, the larger is $\delta$ (hence smaller $b_{\text{old}}$) the larger the climate change is; for $\delta=0$ ($b_{\text{old}}=b_{\text{new}}$) there is no climate change. In section \ref{sec:compar}, we use this interpretation to compare our simulation results to the fossil data. 

We now check how the species survival/extinction may change for a higher evolutionary rate. 
Figure \ref{fig3} shows the critical values of $b_{\text{old}}$ (which, for a given value of $b_{\text{new}}$, is equivalent to the critical value of $\delta$) calculated for different values of $\gamma$. Thus, we readily observe that for a higher rate of adaptive evolution, extinction becomes less likely, i.e.~species survival occurs in a broader parameter range. This is in full agreement with intuitive expectations. 

\begin{figure}[ht]
\centering
\includegraphics[width=0.45\linewidth]{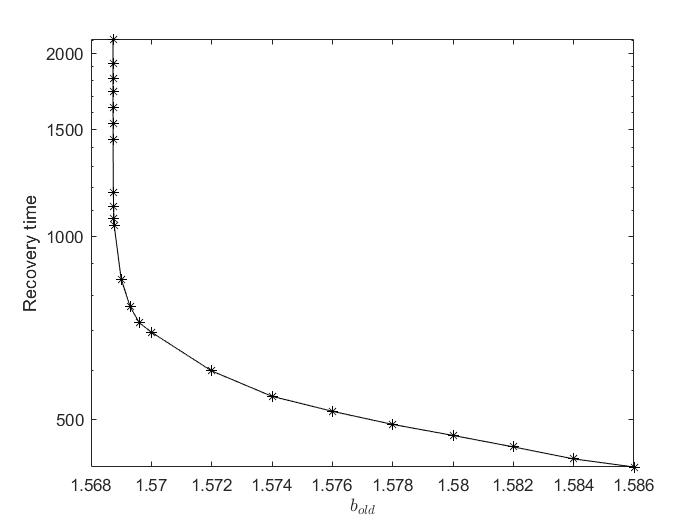}
\includegraphics[width=0.45\linewidth]{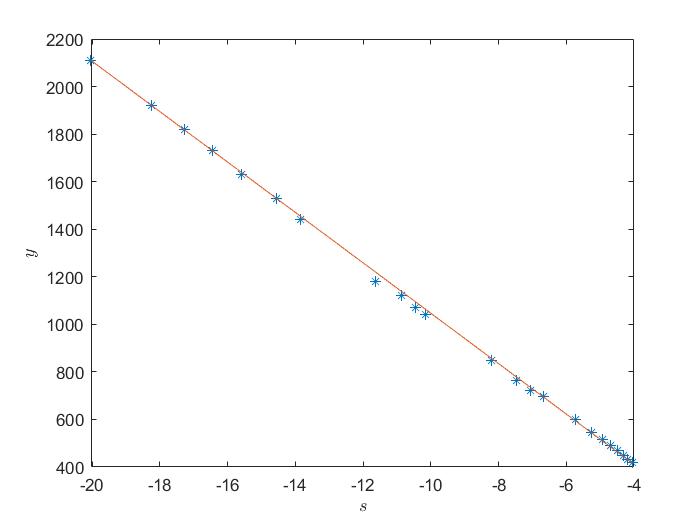}\\
\hspace{9mm} (a) \hspace{55mm} (b)
\caption{{\small (a) Recovery time $y$ (the duration of long transient dynamics) as a function of of $b_{\text{old}}$ for system (\ref{sys2T}-\ref{sys2u},\ \ref{b_1(t)}). Parameter values are given in the text. (b) Same as (a) but shown on the semi-logarithmic scale: $y=y(s)$ where $s=\log(b_{\text{old}}-b_*)$ and $b_*=1.568731028$.}}
\label{fig4} 
\end{figure}

In conclusion of this section, we look into the properties of the long transient dynamics shown in Fig.~\ref{fig2_c}. Specifically, we are interested to know how the duration of the transient - say, $y$ - to which we refer as the recovery time (the time taken by the system dynamics to recover the phytoplankton density from the very small values to which it initially falls) depends on the closeness of parameter $b_{\text{old}}$ to its bifurcation value: the property known in the literature as the scaling law, see \cite{morozov2020} and further references there. 
Figure \ref{fig4}a shows the recovery time obtained for different values of $b_{\text{old}}$. The results indicate that when the value of $b_{\text{old}}$ approaches its critical value, the recovery time grows to infinity; however, the rate of the increase (the scaling law) is unclear. In order to reveal the scaling law, in Fig.~\ref{fig4}b the results are shown on semi-logarithmic scale and are well approximated by a straight line. We therefore arrive at the following scaling law:
\begin{equation}\label{LTscaling}
y = b-a \log(b_{\text{old}}-b_*),
\end{equation}
where $a$ and $b$ are certain coefficients. Thus, the long transient's duration grows rather slowly compared to the exponential and power laws usually observed in nonlinear systems~\cite{morozov2020}. 


\section{Extinctions in multi-species model with differential evolutionary rates}\label{nu_2}

In this section, we attempt to abate one of the main limitations of our model, namely, that the phytoplankton density is described by a single variable while in reality phytoplankton consists of many different species. 
Correspondingly, we now consider an extension of the model where different phytoplankton species are included explicitly. Let phytoplankton community consist of $m$ species, then the state of system is described by $m$ variables, ${\bf u}=(u_1,u_2,\ldots,u_m)$, where $u_k$ is the $k$th species density. As above, we assume that the growth of each species is described by the logistic equation but with different (species-specific) parameters. The species compete for common resources; we assume that the competition affects species per capita growth rates but not carrying capacities. Thus, we arrive at the following model: 
\begin{equation}\label{HX1}
\frac{du_k}{dt}=u_k \left(B_k(T,{\bf R})  -  u_k\right), \quad k=1,\dots, m,
\end{equation}
(cf.~\cite{huisman1999,kozlov2017}) where $B_k$ is the $k$th species's per capita growth rate, 
and ${\bf R}=(R_1,\ldots,R_n)$ are resources, e.g.~nutrients \cite{huisman1999}. 

In this paper, for the sake of simplicity we consider the case where resources are aplenty and hence their availability is not a limiting factor for the species population dynamics. Therefore, the dependence of $B_k$ on ${\bf R}$ in Eqs.~(\ref{HX1}) can be dropped. Instead, we focus on the effect of the temperature, for which we consider the same parametrization as above, see Fig.~\ref{s0s1}b and Eq.~(\ref{plankgrowtemp}), but with some of the coefficients being now species-specific. 
Thus, we arrive at the following multi-species temperature-phytoplankton system:
\begin{align}
\frac{dT}{dt}&=\frac{1}{\lambda}[-aT^4+b(1-A(s_1-s_0)e^{-L({\bf u})}-s_0)], \label{sys2TMod} \\
\frac{du_k}{dt}&=[b_{1,k}(t) e^{-\frac{T_0}{T}}e^{-\alpha_2 T}-u_k]u_k-\sigma u_k, \quad k=1,\ldots,m \label{sys2uMod}
\end{align}
where $L({\bf u})=\sum_{k=1}^m \alpha_{1,k}u_k$. The species can evolve to adapt to the climate changes so that
\begin{equation}\label{multievolv}
b_{1,k}(t) = b_{\text{new}}-(b_{\text{new}}-b_{\text{old},k}) e^{-\gamma_k t},
\end{equation}
where we assume that parameters $b_{\text{old},k}$ and $\gamma_k$ are species-specific but $b_{\text{new}}$ is the same for all species. 

The choice of $b_{\text{old},k}$ and $\gamma_k$ requires a brief discussion. 
These parameters quantify certain biological traits of phytoplankton species. The traits of currently existing species have developed in the course of evolution. Evolution acts through a random selection aiming to maximize the species fitness. Because of this inherent randomness, a particular biological parameter can often be regarded as a random number drawn from a certain probability distribution \cite{shipley2006}. 
This approach is particularly relevant in the case of a community consisting of many similar species \cite{carmona2019}. 
Correspondingly, for the values of $b_{\text{old},k}$ and $\gamma_k$, we consider them being randomly distributed, that is, the arrays $\{b_{\text{old},k},\ k=1,\ldots,m\}$ and $\{\gamma_k,\ k=1,\ldots,m\}$ both consist of $m$ random numbers. 
One can expect that the properties of the multi-species system (\ref{sys2TMod}-\ref{sys2uMod}), in particular the fraction of species that goes extinct in response to a sudden climate change, depend on the properties of the probability distributions. As we will show it below, this is indeed the case.  

\begin{figure}[!t]
  \centering
    \includegraphics[width = 0.32\linewidth]{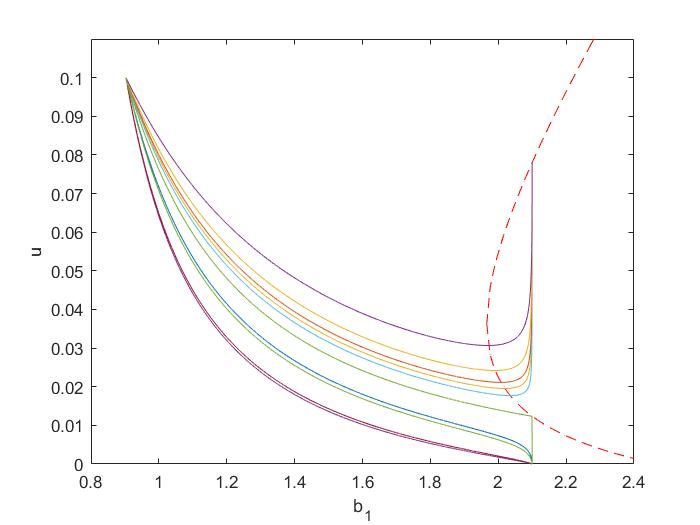}
    \includegraphics[width = 0.32\linewidth]{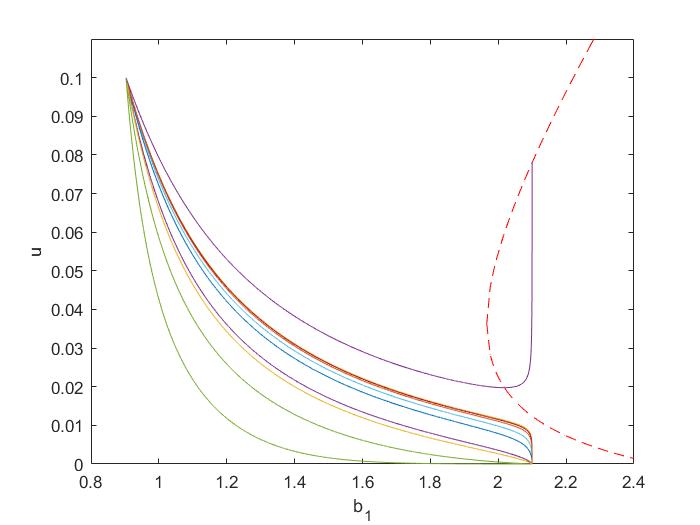}
    \includegraphics[width = 0.32\linewidth]{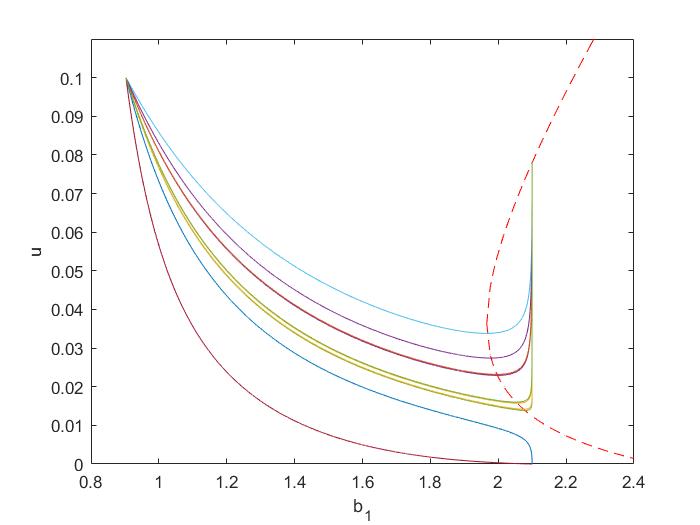}\\
\hspace{9mm} (a) \hspace{45mm} (b) \hspace{45mm} (c)
\caption{{\small The ten colored curves show ten system's trajectories in $(b_1,u)$ plane obtained for ten different values of $\gamma$ randomly chosen from the normal random distribution with mean=0.03 and variance=0.0001. Panels (a,b,c) show three simulation runs. Here $b_{\text{old}}=0.903$, other parameters are the same as above. The dashed red curve is the steady state values of $u$ for the baseline temperature-phytoplankton system (\ref{sys2T}-\ref{sys2u}) as a function of the controlling parameter $b_1$. }}
      \label{f_}
\end{figure}

We begin with the example where $b_{\text{old},k}$ are the same for all species but $\gamma_k$ are described by a normal distribution. 
Figure \ref{f_} shows the results obtained for a community of $m=10$ species. The coloured curves visualize the ten corresponding trajectories in the $(b_1,u)$ plane obtained for $b_{\text{old}}=0.903$ and ten different random values of $\gamma_k$.
Simulations are repeated three times, as shown in panels (a), (b) and (c). Since in each simulation the array of ten random values for $\gamma_k$ is different, the results are different too. In particular, the fraction of species that goes extinct changes between the simulation runs. In Fig.~\ref{f_}a, five out of ten species go extinct while in Figs.~\ref{f_}b and \ref{f_}c it is, respectively, nine and two species go extinct. 

Thus, by repeating the above simulations sufficiently many times, one can obtain a distribution of extinction frequencies. That is, by repeating the simulations $N$ times, in $N_1$ simulation runs only one species goes extinct, in $N_2$ runs two species go extinct, etc., with $N_1+N_2+\ldots+N_m=N$. 

\begin{figure}[!t]
\centering
 \includegraphics[width=0.45\linewidth]{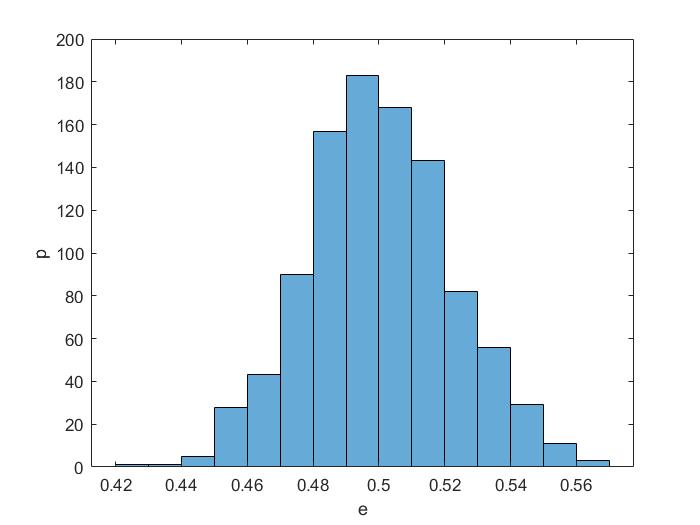}
 \includegraphics[width=0.45\linewidth]{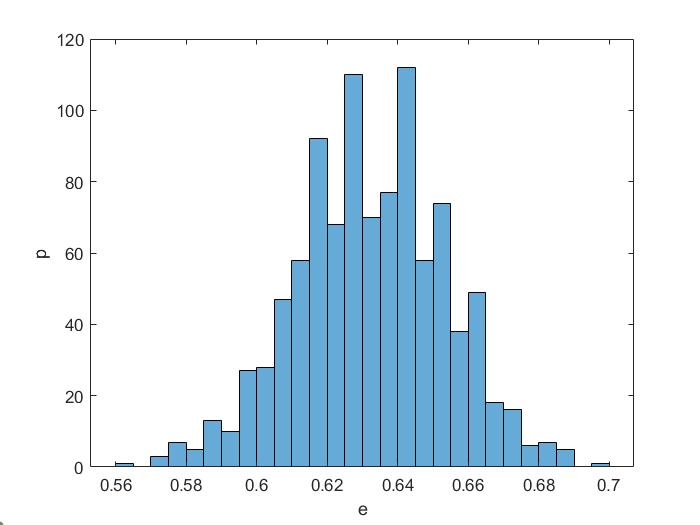}\\
\hspace{9mm} (a) \hspace{55mm} (b)
\caption{{\small Frequencies $p$ of the number of extinctions $e$ (shown as a fraction of the total number of species, hence $0\le e\le 1$) obtained from system (\ref{sys2TMod}-\ref{sys2uMod}) with $m=500$ species. Simulations are repeated $N=1000$ times. (a) Values of $\gamma_k$ are normally distributed with mean=0.03 and variance=0.0001; (b) values of $\gamma_k$ are exponentially distributed with mean=0.03. 
Here $b_{\text{old}}=0.903$, other parameters are given in the text.}}
\label{dis_2} 
\end{figure}

Figure \ref{dis_2}a shows the distribution of extinction frequencies obtained using the above approach with the number of species $m=500$ (with $\gamma_k$ being normally distributed) and the total number of simulation runs $N=1000$. The distribution of extinction frequencies appears to be bell-shaped, with the position of the maximum at $e\approx 0.5$. However, this shape is in stark difference with the fossil record that shows that there are only a few large extinctions ($e>0.5$), cf.~the ``Big Five'', but many extinctions of a much smaller magnitude ($0.05<e<0.2$) \cite{wignall2019}. Therefore, the extinctions frequency should be a monotonously decreasing curve with the maximum at or close to $e=0$ rather than a bell-shaped curve. Thus, something is amiss in the model. 

In order to understand how the agreement between the simulation results and the data could be improved, we first consider a different distribution of $\gamma_k$. Figure \ref{dis_2}b shows the distribution of extinction frequencies obtained for $\gamma_k$ being exponentially distributed. We readily observe that the shape of the distribution does not change much, remaining bell-shaped, with only a slightly different position of the maximum. 

\begin{figure}[!b]
\vspace*{3mm}
  \centering
  \begin{subfigure}{.3\linewidth}
    \includegraphics[width = \linewidth]{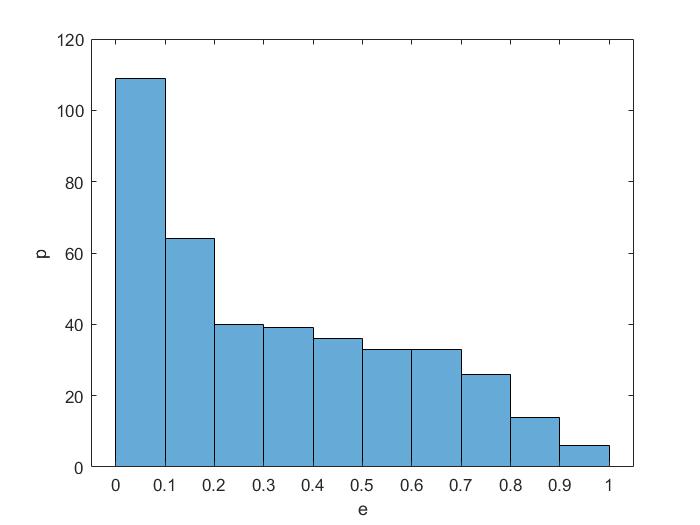}
    \caption{}
    \label{dis_8_a}
  \end{subfigure}%
  \hspace{1em}
  \begin{subfigure}{.3\linewidth}
    \includegraphics[width = \linewidth]{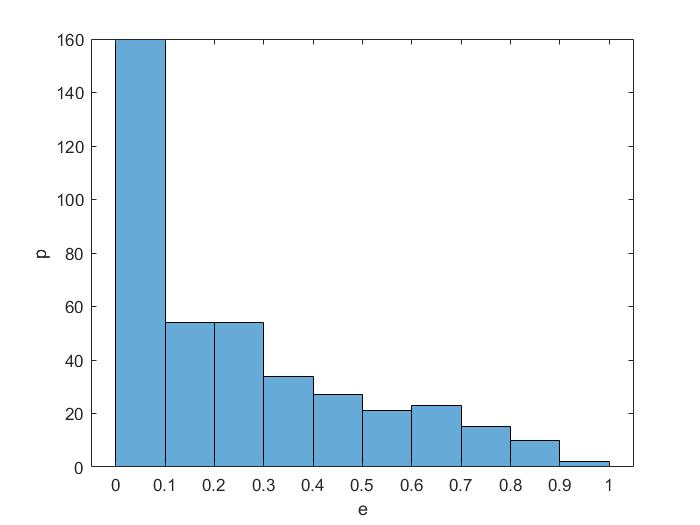}
    \caption{}
    \label{dis_8_b}
  \end{subfigure}%
  \hspace{1em}
  \begin{subfigure}{.3\linewidth}
    \includegraphics[width = \linewidth]{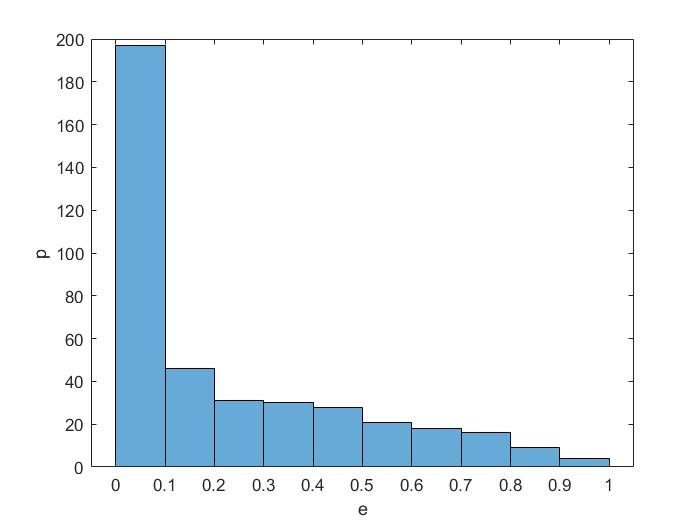}
    \caption{}
    \label{dis_8_c}
  \end{subfigure}
\caption{{\small Probability density function ($p$) of the number of extinctions ($e$) (shown as a fraction of the whole number of species) for temperature-phytoplankton system with different values of $b_{\text{old}}$ (a sequence of 400 elements) obtained from exponential random distribution (calculated as $b_{\text{old}}=\bar{b}-h$, where $\bar{b}=2, \ 2.1, \ {\text{and}}\ 2.2$ in figures \ref{dis_8_a}, \ref{dis_8_b} and \ref{dis_8_c}, respectively, and $h$ is exponentially distributed, i.e. with the probability distribution given as $p(h_1)= we^{-w h}$ with $w=1.8$), and different values of $\gamma$ (a sequence of 100 elements) obtained from exponential random distribution. Other parameters are given in the text. Note that, since $b_{\text{old}}$ can be regarded as a proxy for a climate change event, the size of the array $\{b_{\text{old},k}\}$ can be different from $\{\gamma_k\}$.}}
  \label{dis_8}
\end{figure}

A key to resolving this difficulty (i.e.~the apparent disagreement with fossil record) lies in the understanding that, generally speaking, some other parameters of the model should also be regarded as random values. Specifically, we now proceed to a more general case where $b_{\text{old}}$ is also random. 
Figure \ref{dis_8} shows the distribution of extinction frequencies obtained in system (\ref{sys2TMod}-\ref{sys2uMod}) for the array $\{b_{\text{old},k},\ k=1,\ldots,m\}$ consisting of 400 random exponentially distributed numbers. More specifically, we assume that the probability of a climate change decreases with its magnitude. Since the magnitude of the change in our model is quantified by the distance between the bifurcation curve and $b_{\text{old}}$, we calculate it as $b_{\text{old}}=\bar{b}-h$ where 
$h$ is exponentially distributed with the probability density given as $p(h)= we^{-w h}$, where $\bar{b}$ and $w$ are parameters. 
Figures \ref{dis_8_a}, \ref{dis_8_b} and \ref{dis_8_c} shows the results obtained for $w= 1.8$ and $\bar{b}=2, \ 2.1, \ {\text{and}}\ 2.2$, respectively. 
In all three cases, $m=100$ values of $\gamma$ are distributed exponentially with mean=0.03. 
Thus, changing constant parameter $b_{\text{old}}$ to an array of random numbers changes the results dramatically. Instead of a bell-shaped distribution of extinction frequencies (as shown in Fig.~\ref{dis_2}), we know obtain a monotonously decreasing distribution, which is in a qualitative agreement with the fossil record \cite{wignall2019}, see also below.

\subsection{Comparison between the simulations and the data}\label{sec:compar}

\begin{figure}[!b]
\centering
 \includegraphics[width=0.6\linewidth]{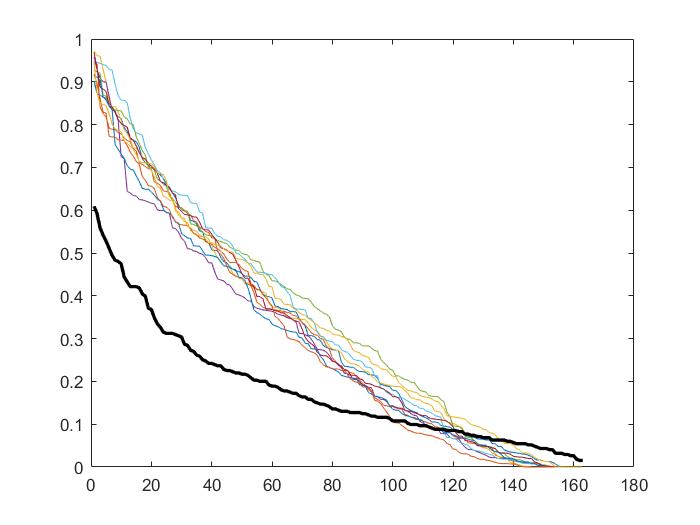}\\
\caption{{\small The same as figure \ref{dis_8_a} but for 163 values of $b_{\text{old}}$. The ten coloured curves corresponds to ten simulation runs using parameter values the same as in Fig.~\ref{dis_8_a}. The thick black curve refers to the historical data.}}
\label{dis_10_data} 
\end{figure}

In this section, we endeavour to compare our simulation results to historical data on extinctions in a more explicit way. Fossil record contains data on 163 extinction events (including the ``Big Five'') with the extinction magnitude, i.e.~the estimated percentage of all species that went extinct, ranging approximately from 5\% to 68\%. 
The thick black curve in Fig.~\ref{dis_10_data} shows these extinctions events in the ranked order, i.e.~from the largest to the smallest. 

We now recall that different values of $b_{\text{old}}$ can be regarded as a climate change of different magnitude. Correspondingly, in order to provide a direct comparison between the simulation results and the data, we now repeat the simulations with 163 different values of $b_{\text{old}}$ and show the obtained fractions of extinct species in the ranked order, i.e.~from largest to smallest. The results are shown in Fig.~\ref{dis_10_data}. 
To account for the system randomness, the simulations are repeated ten times; it appears that the random fluctuations have a relatively little effect on the results. 

We readily observe that, although our results are in a general agreement with the data, e.g.~predicting mass extinctions of the magnitude comparable with those shown by the fossil record, there are apparent disagreements too. In particular, our results overestimate the frequency of large-scale extinctions (with more than 60\% of species going extinct) and slightly under-estimate the frequency of small extinctions. Possible reasons for this disagreement are discussed in the next section.

\section{Discussion and conclusions}\label{sec:discuss}

Mass extinctions were an important part of the history of life on Earth \cite{bambach2006,sepkoski1986,wignall2019}. 
It is widely believed that the ultimate reason for a mass extinction is a sufficiently large climate change \cite{bond2017,sudakow2022}. This, in turn, can be caused by a sufficiently large (over-critical) perturbation of CO$_2$ cycle \cite{rothman2017}. A large climate change can
eventually lead to species extinctions and considerable biodiversity loss on the global scale through a variety of specific mechanisms or pathways, the most common one being a significant change in the average Earth temperature resulting in global warming or global cooling. 

While there has been a great progress over the last two decades in understanding the main causes and triggers leading to a mass extinctions, e.g.~see \cite{alroy2008,bambach2006,bond2017,harnik2012,melott2014,rothman2017,rothman2019a,wignall2019}, many questions remain open. In particular, it is well known that not every climate change in the Earth history resulted in mass extinction; thus, there must exist factors or feedbacks that can attenuate the effect of the change. Due to the intrinsic deficiencies of paleontological data (for details and a discussion of this issue see Section 3 in~\cite{sudakow2022}), it does not seem possible to identify such factors or feedbacks based solely on the analysis of fossil record. Mathematical models are needed; mechanistic process-based models create a ‘virtual laboratory’ where specific hypotheses can be tested and various scenarios investigated. Mathematical modelling has long become a part of a standard toolbox in ecology as well as environmental sciences more generally; surprisingly, application of models in paleontology is still rare. 

In this paper, we have developed a new mathematical model with a potential to describe a mass extinction using the phytoplankton community as a proxy to broader taxa. 
In particular, our model reveals how the extinction magnitude can be affected by the climate-vegetation coupling (through phytoplankton active feedback to the climate) and by the species evolutionary response to the climate change: the two factors that are apparently important yet have been largely overlooked in the existing literature. We have shown that our model (see Eqs.~(\ref{sys2TMod}-\ref{multievolv})) predicts the distribution of extinction frequencies (e.g.~as in Figs.~\ref{dis_8} and \ref{dis_10_data}) that is consistent with the data. 

In spite of the general consistency, there are some apparent differences between the ranked order of extinctions predicted by our model and that shown by the fossil record. The simulated distribution is described by a steeper curve, hence overestimating the frequency of large extinctions and underestimating the frequency of small ones. However, we believe that this is not an inherent flaw of the model and the agreement with the data can, in principle, be improved. In particular, here we recall that the results depend significantly on the type of parameter $b_{\text{old}}$. Considering it as an exponentially distributed random value (instead of a fixed single value) leads to a qualitative change in the extinctions frequency distribution changing it from a bell-shaped distribution to a monotonous one, cf.~Figs.~\ref{dis_2} and \ref{dis_8}. 
Thus, one can expect that a different choice of the probability distribution for $b_{\text{old}}$ (and perhaps for $\gamma$) may cause further changes in the extinctions frequency distribution making it closer to the data. This should become a focus of future work. 

At the same time, we also want to mention that the fossil record gives only, at best, a partial view of the actual extinction magnitude. The matter is that the fossil record contains predominantly data on hard-bodied species (e.g.~mollusks and vertebrates) while soft bodied species (to which most of phytoplankton species belong) usually disappear without leaving any trace. How the distribution of extinctions (cf.~thick black curve in Fig.~\ref{dis_10_data}) may change if data on soft-bodied species are included is an entirely open issue. This may partially explain the disagreement between our model and the fossil record. A model allowing for a direct comparison with the fossil record has to include explicitly a hard-bodied taxa, e.g.~some zooplankton species. That should become another focus of future work.


\vspace*{7mm}




\clearpage

\appendix

\section{Checking results robustness to parameter values}

Since most of the results in the main text were obtained numerically, the question arises as to how a change in parameter values may change the system's properties. We are mostly interested to the effect of variations in $\gamma$, as this parameter describes the rate of species's adaptive evolution. Correspondingly, below we repeat some of the simulations of Section \ref{sec:evolut1sp}, specifically those shown in Figs.~\ref{fig2}, \ref{traj} and \ref{fig4}, for two different values of $\gamma$.

\subsection{$\gamma=0.03$} 

Here we repeat the simulations shown in Figs.~\ref{fig2}, \ref{traj} and \ref{fig4} with a larger value of $\gamma$, $\gamma=0.03$. The results are shown below, respectively, in Figs.~\ref{ev3}, \ref{traj_3} and \ref{scale_3}.

\begin{figure}[!b]
  \centering
  \begin{subfigure}{.4\linewidth}
    \includegraphics[width = \linewidth]{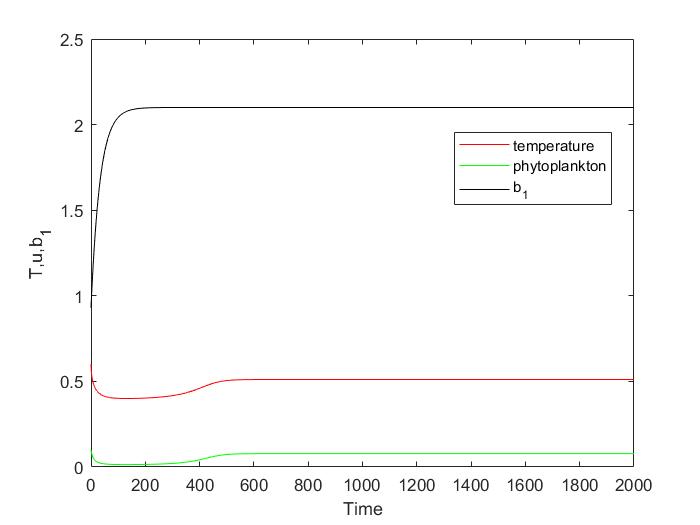}
    \caption{}
    \label{ev3_a}
  \end{subfigure}%
  \hspace{0em}
  \begin{subfigure}{.4\linewidth}
    \includegraphics[width = \linewidth]{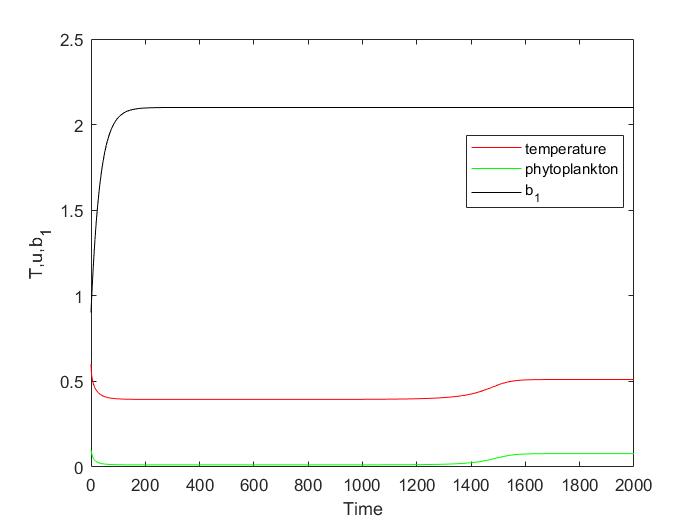}
    \caption{}
    \label{ev3_c}
  \end{subfigure}
    \hspace{0em}
  \begin{subfigure}{.4\linewidth}
    \includegraphics[width = \linewidth]{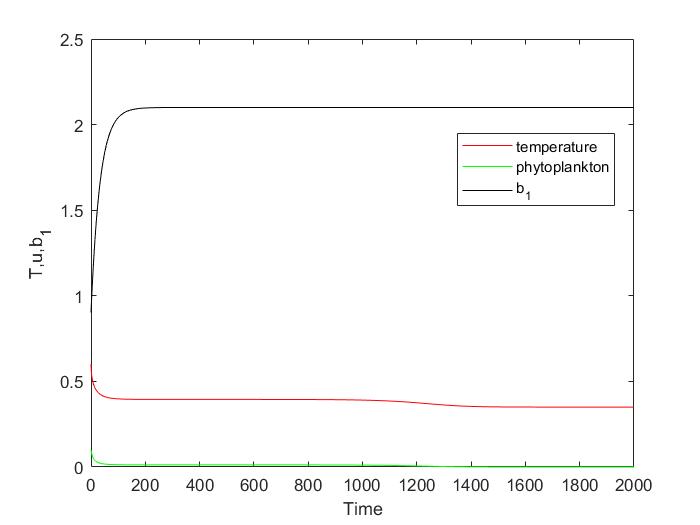}
    \caption{}
    \label{ev3_d}
  \end{subfigure}%
  \begin{subfigure}{.4\linewidth}
    \includegraphics[width = \linewidth]{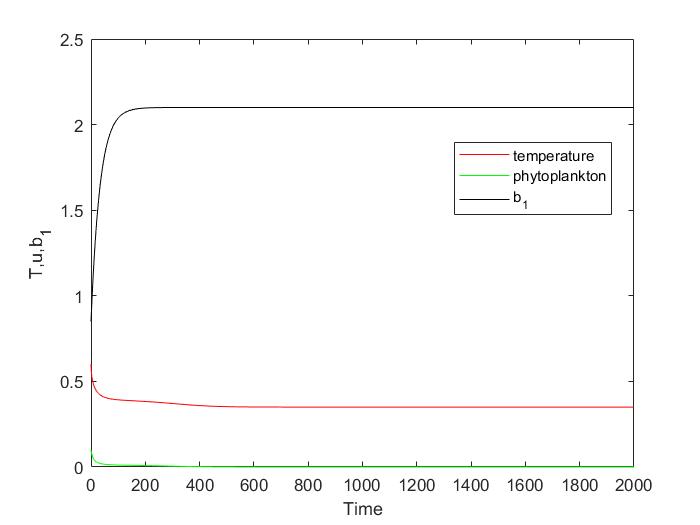}
    \caption{}
    \label{ev3_e}
  \end{subfigure}%
\caption{{\small Averaged temperature (red) and phytoplankton density (green) vs time as given by system (\ref{sys2T}-\ref{sys2u}) with $b_1(t)$ (black) defined by Eq.~(\ref{b_1(t)}) for different values of parameter $b_{\text{old}}$: (a) $b_{\text{old}}=0.93$, (b) $b_{\text{old}}=0.90322$, (c) $b_{\text{old}}=0.90321$ and (d) $b_{\text{old}}=0.85$.}}
  \label{ev3}
\end{figure}
\begin{figure}[ht]
\centering
\includegraphics[width=0.6\linewidth]{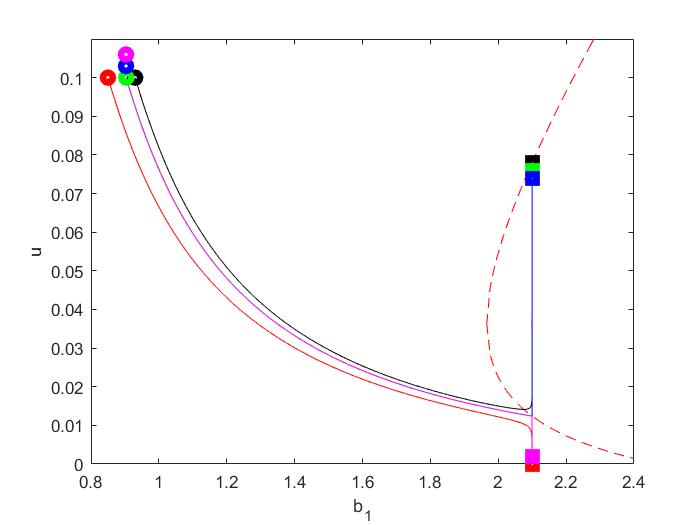}
\caption{{\small Dynamics of the system (\ref{sys2T}-\ref{sys2u},\ \ref{b_1(t)}) shown in the $(b_1,u)$ plane for five different values of $b_{\text{old}}$: $b_{\text{old}}=0.93$, 0.904, 0.90322,0.90321 and 0.85 (black, green, blue, magenta and red colours, respectively). The circles are the start points and the squares are the end points. The dashed red curve shows the steady state values of $u$ (the bifurcation curve) in the baseline temperature-phytoplankton system without evolution, cf.~Eqs.~ (\ref{sys2T}-\ref{sys2u}).}}
\label{traj_3} 
\end{figure}
\begin{figure}[ht]
\centering
\includegraphics[width=0.45\linewidth]{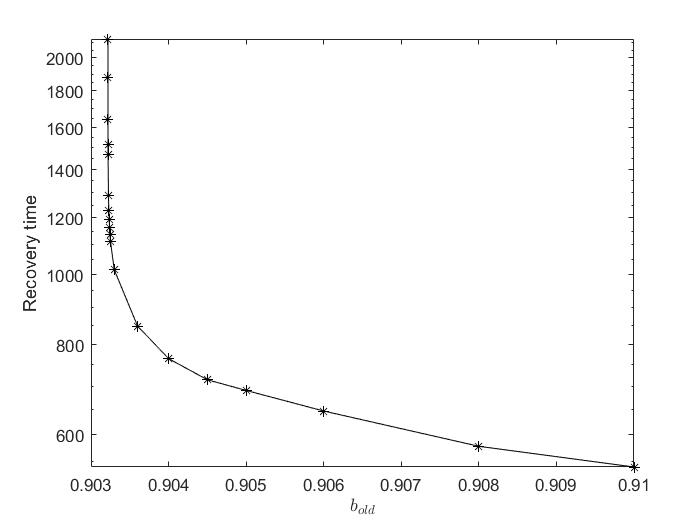}
\includegraphics[width=0.45\linewidth]{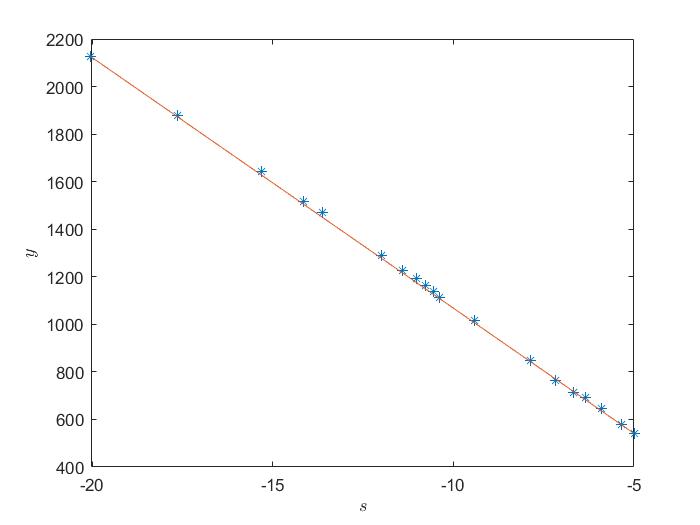}
\caption{{\small (a) Recovery time $y$ (the duration of long transient dynamics) as a function of of $b_{\text{old}}$ for system (\ref{sys2T}-\ref{sys2u},\ \ref{b_1(t)}). (b) Same as (a) but shown on the logarithmic scale: $y=y(s)$ where $s=\log z$, $z=b_{\text{old}}-b_*$ and $b_*=0.903218778$.}}
\label{scale_3} 
\end{figure}

\clearpage

\subsection{$\gamma=0.05$}

Now, we repeat the simulations shown in Figs.~\ref{fig2}, \ref{traj} and \ref{fig4} with $\gamma=0.05$. The results are shown below, respectively, in Figs.~\ref{ev4}, \ref{traj_4} and \ref{scale_4}. We readily observe that the results are qualitatively the same, with only slightly different numerical values. 

\begin{figure}[!h]
\vspace*{5mm}
  \centering
  \begin{subfigure}{.4\linewidth}
    \includegraphics[width = \linewidth]{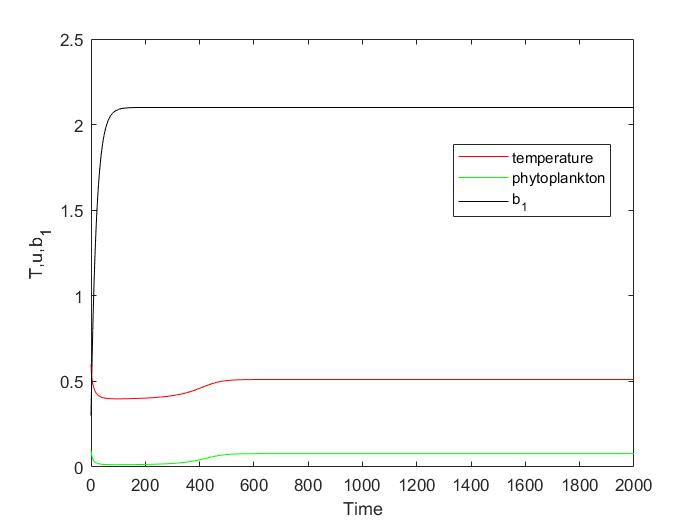}
    \caption{}
    \label{ev4_a}
  \end{subfigure}%
  \hspace{0em}
  \begin{subfigure}{.4\linewidth}
    \includegraphics[width = \linewidth]{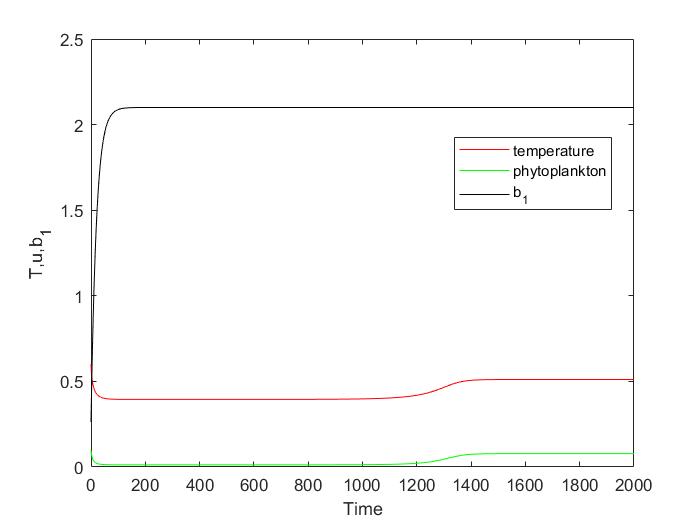}
    \caption{}
    \label{ev4_c}
  \end{subfigure}
    \hspace{0em}
  \begin{subfigure}{.4\linewidth}
    \includegraphics[width = \linewidth]{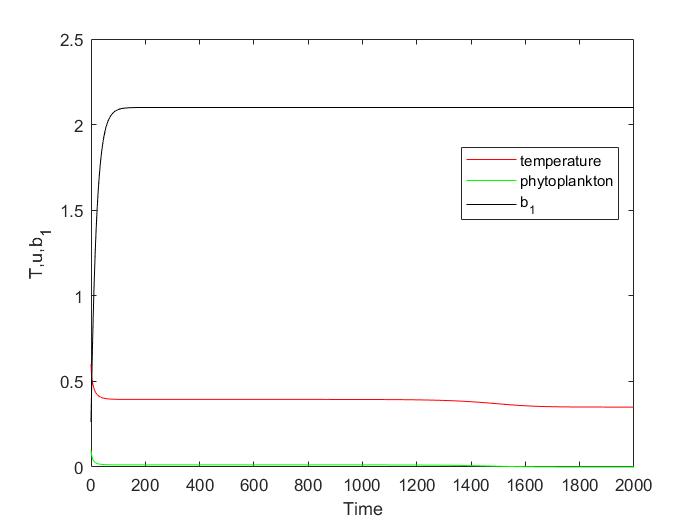}
    \caption{}
    \label{ev4_d}
  \end{subfigure}%
  \begin{subfigure}{.4\linewidth}
    \includegraphics[width = \linewidth]{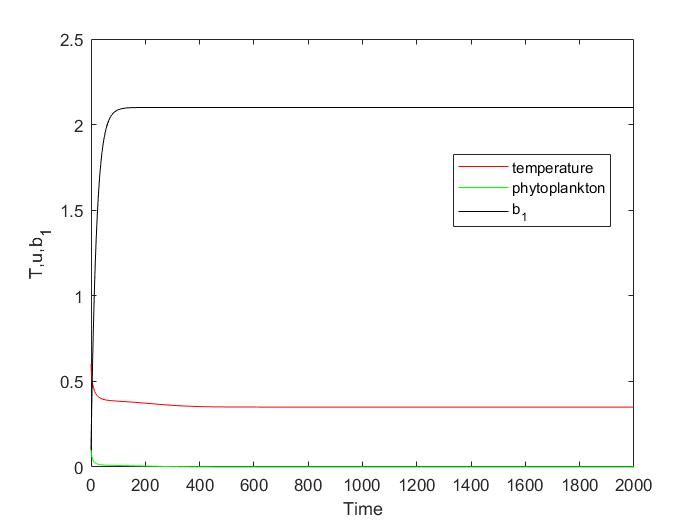}
    \caption{}
    \label{ev4_e}
  \end{subfigure}%
\caption{{\small Averaged temperature (red) and phytoplankton density (green) vs time as given by system (\ref{sys2T}-\ref{sys2u}) with $b_1(t)$ (black) defined by Eq.~(\ref{b_1(t)}) for different values of parameter $b_{\text{old}}$: (a) $b_{\text{old}}=0.3$, (b) $b_{\text{old}}=0.26307$, (c) $b_{\text{old}}=0.26306$ and (d) $b_{\text{old}}=0.1$.}}
  \label{ev4}
\end{figure}

\begin{figure}[ht]
\centering
\includegraphics[width=0.6\linewidth]{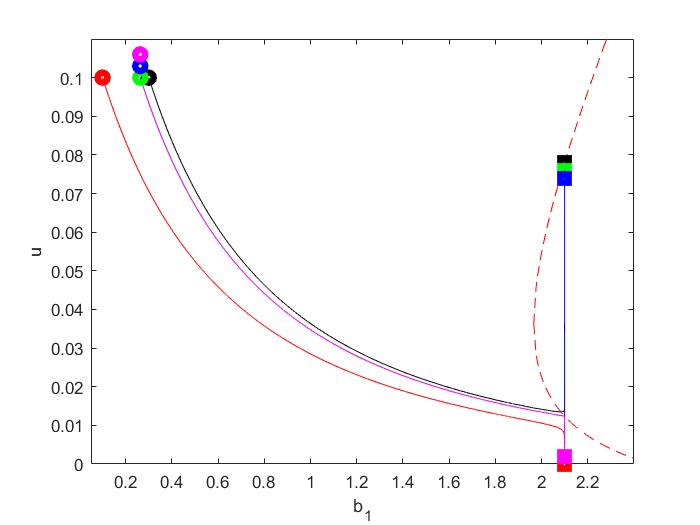}
\caption{{\small Dynamics of the system (\ref{sys2T}-\ref{sys2u},\ \ref{b_1(t)}) shown in the $(b_1,u)$ plane for five different values of $b_{\text{old}}$: $b_{\text{old}}=0.3$, 0.264, 0.26307,0.26306 and 0.1
(black, green, blue, magenta and red colours, respectively). The circles are the start points and the squares are the end points. The dashed red curve shows the steady state values of $u$ (the bifurcation curve) in the baseline temperature-phytoplankton system without evolution, cf.~Eqs.~ (\ref{sys2T}-\ref{sys2u}).}}
\label{traj_4} 
\end{figure}
\begin{figure}[ht]
\centering
\includegraphics[width=0.45\linewidth]{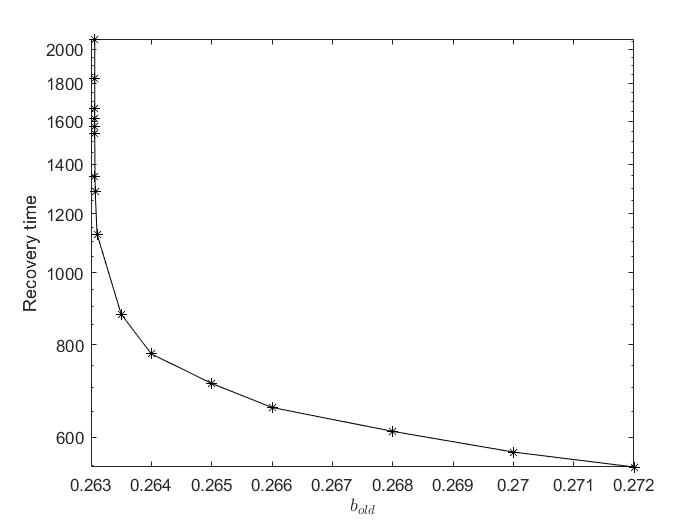}
\includegraphics[width=0.45\linewidth]{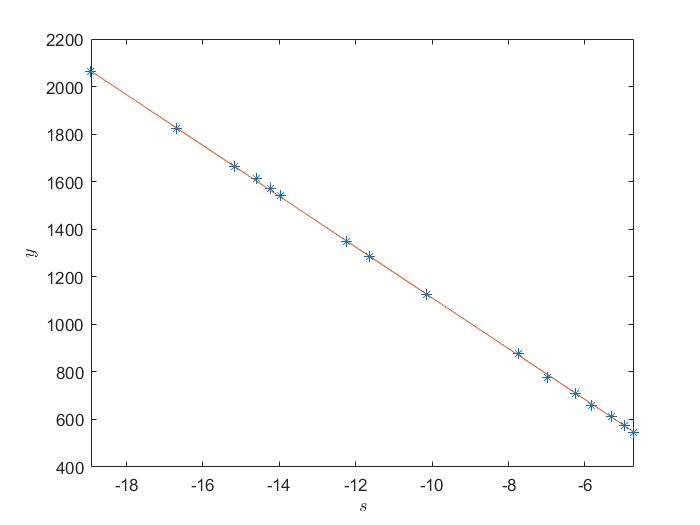}
\caption{{\small (a) Recovery time $y$ (the duration of long transient dynamics) as a function of of $b_{\text{old}}$ for system (\ref{sys2T}-\ref{sys2u},\ \ref{b_1(t)}). (b) Same as (a) but shown on the logarithmic scale: $y=y(s)$ where $s=\log z$, $z=b_{\text{old}}-b_*$ and $b_*=0.263061144$.}}
\label{scale_4} 
\end{figure}

\end{document}